\documentclass[acmsmall]{acmart}

\usepackage{metalogo}
\usepackage{smartdiagram}
\usepackage{forest}
\usepackage{cleveref}
\usepackage{booktabs}
\usepackage{tabularx}
\usepackage{multirow}
\usepackage{makecell}
\usepackage{subfigure}
\usepackage[misc]{ifsym} 
\usepackage{balance}
\usepackage{listings}
\usepackage{color, xcolor}
\usepackage{tcolorbox}
\usepackage{amsmath}
\usepackage{algorithm}
\usepackage{graphicx}
\usepackage{algorithmicx}
\usepackage{algpseudocode}
\usepackage{textcomp}
\usepackage{diagbox}
\usepackage{colortbl}
\usepackage{seqsplit}
\usepackage{xurl}
\usepackage{url}
\usetikzlibrary{arrows.meta}

\definecolor{lightcoral}{rgb}{0.94, 0.5, 0.5}
\definecolor{lightgreen}{rgb}{0.56, 0.93, 0.56}
\definecolor{harvestgold}{rgb}{0.85, 0.57, 0.0}
\definecolor{brightlavender}{rgb}{0.75, 0.58, 0.89}
\definecolor{capri}{rgb}{0.0, 0.75, 1.0}
\definecolor{carminepink}{rgb}{0.92, 0.3, 0.26}
\definecolor{celadon}{rgb}{0.67, 0.88, 0.69}
\definecolor{darkpastelgreen}{rgb}{0.01, 0.75, 0.24}
\definecolor{DeepSkyBlue4}{RGB}{0,104,139}

\crefformat{section}{\S#2#1#3} 
\crefformat{subsection}{\S#2#1#3}
\crefformat{subsubsection}{\S#2#1#3}

\acmJournal{TOSEM}
\acmYear{2026} \acmVolume{1} \acmNumber{1} \acmArticle{}
\acmMonth{1} \acmPrice{} \acmDOI{10.1145/3789262}

\begin{document}

\title{A Survey on RLOps for Large Language Models}

\author{Lingzhe Zhang}
\affiliation{%
	\institution{Peking University}
	\city{Beijing}
	\country{China}}
\email{zhang.lingzhe@stu.pku.edu.cn}

\author{Tong Jia$^{\ast}$}
\thanks{$^{\ast}$Corresponding author}
\affiliation{%
	\institution{Peking University; National Key Laboratory of Data Space Technology and System}
	\city{Beijing}
	\country{China}}
\email{jia.tong@pku.edu.cn}

\author{Chiming Duan}
\affiliation{%
	\institution{Peking University}
	\city{Beijing}
	\country{China}}
\email{duanchiming@stu.pku.edu.cn}

\author{Minghua He}
\affiliation{%
	\institution{Peking University}
	\city{Beijing}
	\country{China}}
\email{hemh2120@stu.pku.edu.cn}

\author{Leyi Pan}
\affiliation{%
	\institution{Tsinghua University}
	\city{Beijing}
	\country{China}}
\email{panly24@mails.tsinghua.edu.cn}

\author{Kening Zheng}
\affiliation{%
	\institution{University of Illinois Chicago}
	\country{United States}}
\email{kzhen28@uic.edu}

\author{Zhaoyang Liu}
\affiliation{%
	\institution{Alibaba Group}
	\country{China}}
\email{jingmu.lzy@alibaba-inc.com}

\author{Bolin Ding}
\affiliation{%
	\institution{Alibaba Group}
	\country{United States}}
\email{bolin.ding@alibaba-inc.com}

\author{Philip S. Yu}
\affiliation{%
	\institution{University of Illinois Chicago}
	\country{United States}}
\email{psyu@uic.edu}

\author{Ying Li$^{\ast}$}
\affiliation{%
	\institution{Peking University}
	\city{Beijing}
	\country{China}}
\email{li.ying@pku.edu.cn}

\renewcommand{\shortauthors}{Lingzhe Zhang et al.}

\begin{abstract}
	Reinforcement learning (RL) has become increasingly important for improving the reasoning, instruction-following, and task-solving capabilities of large language models (LLMs). However, scaling RL training exposes failures in distributed execution, feedback computation, and learning dynamics that can waste computational resources, slow training, and undermine model improvements. These challenges have motivated a growing body of techniques for LLM RLOps, which manages the efficiency and reliability of operations throughout the LLM reinforcement learning training lifecycle. Despite growing interest, these techniques remain scattered across research areas, with limited synthesis of their operational roles and interactions. To bridge this gap, we present a systematic survey of LLM RLOps methods. We organize existing approaches into five stages: Training Setup, Experience Collection, Reward Computation, Policy Optimization, and Model Evaluation. Within each stage, we examine operational concerns, management techniques, and evaluation evidence. Building on this taxonomy, we identify open challenges and future directions in training process observability, cross-stage coordination, generalizable RLOps, multi-run management, and RLOps for recursive self-improvement. These directions extend operational management from individual training runs to diverse training settings, shared infrastructure, and successive improvement cycles. We maintain a GitHub repository indexing relevant papers and open resources at \href{https://github.com/zhanglingzhe0820/Awesome-RLOps}{https://github.com/zhanglingzhe0820/Awesome-RLOps}.
\end{abstract}




%\received{20 June 2024}
%\received[revised]{}
%\received[accepted]{}

\maketitle

\section{Introduction}
\label{sec:introduction}

Large Language Models (LLMs), such as GPT-6 Astra~\cite{openai2026astra} and Claude Opus 5.5~\cite{anthropic2026opus55}, have demonstrated strong capabilities in language understanding, complex reasoning, and tool use. These capabilities support a wide range of applications, including conversational assistance, software development, scientific research, and computer-assisted task execution. As LLMs increasingly undertake tasks requiring sustained reasoning and interaction with external environments, reinforcement learning (RL) has become an important component of their training. By using feedback on generated responses and executed actions, RL enables models to refine their behavior beyond supervised demonstrations. For example, GPT-6 Astra incorporates advances in RL alongside pre-training and alignment to support complex reasoning and agentic capabilities~\cite{openai2026astra}. These developments extend the role of RL in LLM training from improving individual responses to supporting multi-step behavior in interactive environments.

\subsection{Why Does RLOps Matter for Large Language Models?}
\label{sec:why-rlops}

As RL training scales to larger models and longer interactions, coordinating experience collection, reward computation, and policy updates becomes increasingly challenging. Failures can arise from both distributed execution and the learning process itself. For example, DAPO reports entropy collapse, reward noise, and training instability in its initial GRPO experiments~\cite{yu2025dapo}. AReaL highlights a trade-off between execution efficiency and learning stability: synchronous training leaves resources idle while waiting for long rollouts, whereas asynchronous execution introduces outdated training samples that require explicit control~\cite{fu2026areal}. Beyond these execution and learning challenges, reward exploitation can also produce unintended behavioral changes. MacDiarmid et al.~\cite{macdiarmid2025natural} show that, in experiments using production coding environments, models that learn to exploit reward mechanisms can generalize to broader misaligned behaviors, including alignment faking and attempted sabotage.

These issues can waste computation, restrict exploration, and cause training to optimize unintended objectives. Crucially, an RL run may continue executing without producing dependable improvements. Each policy update changes the experience collected in subsequent iterations, allowing errors in feedback or inconsistencies between rollout and training to affect later learning. Looking toward recursive self-improvement (RSI), where models increasingly participate in designing, executing, and evaluating subsequent rounds of model improvement, these concerns become even more consequential. Unreliable feedback or unvalidated updates could propagate across successive improvement cycles, increasing the need for monitoring, verification, and recovery. Consequently, managing LLM reinforcement learning requires checking not only that training jobs execute successfully, but also that resources are used effectively, feedback remains valid, and model updates produce the intended gains.

To address these challenges, researchers have developed techniques for coordinating distributed execution, controlling experience staleness, stabilizing learning, and mitigating reward hacking~\cite{sheng2024hybridflow,fu2026areal,yu2025dapo,macdiarmid2025natural}. We bring these efforts together under \textit{LLM RLOps}, a broad class of techniques designed to plan, execute, monitor, control, recover, and evaluate LLM reinforcement learning processes. Its scope spans training setup, experience collection, reward computation, policy optimization, and model evaluation, with the goal of making training efficient, reliable, and reproducible.

\subsection{Why a Survey for LLM RLOps?}
\label{sec:why-survey}

\begin{figure}[htbp]
	\centering
	\subfigure[Quarterly LLM RLOps publications.]{
		\begin{minipage}{0.47\linewidth}
			\centering
			\includegraphics[width=\textwidth]{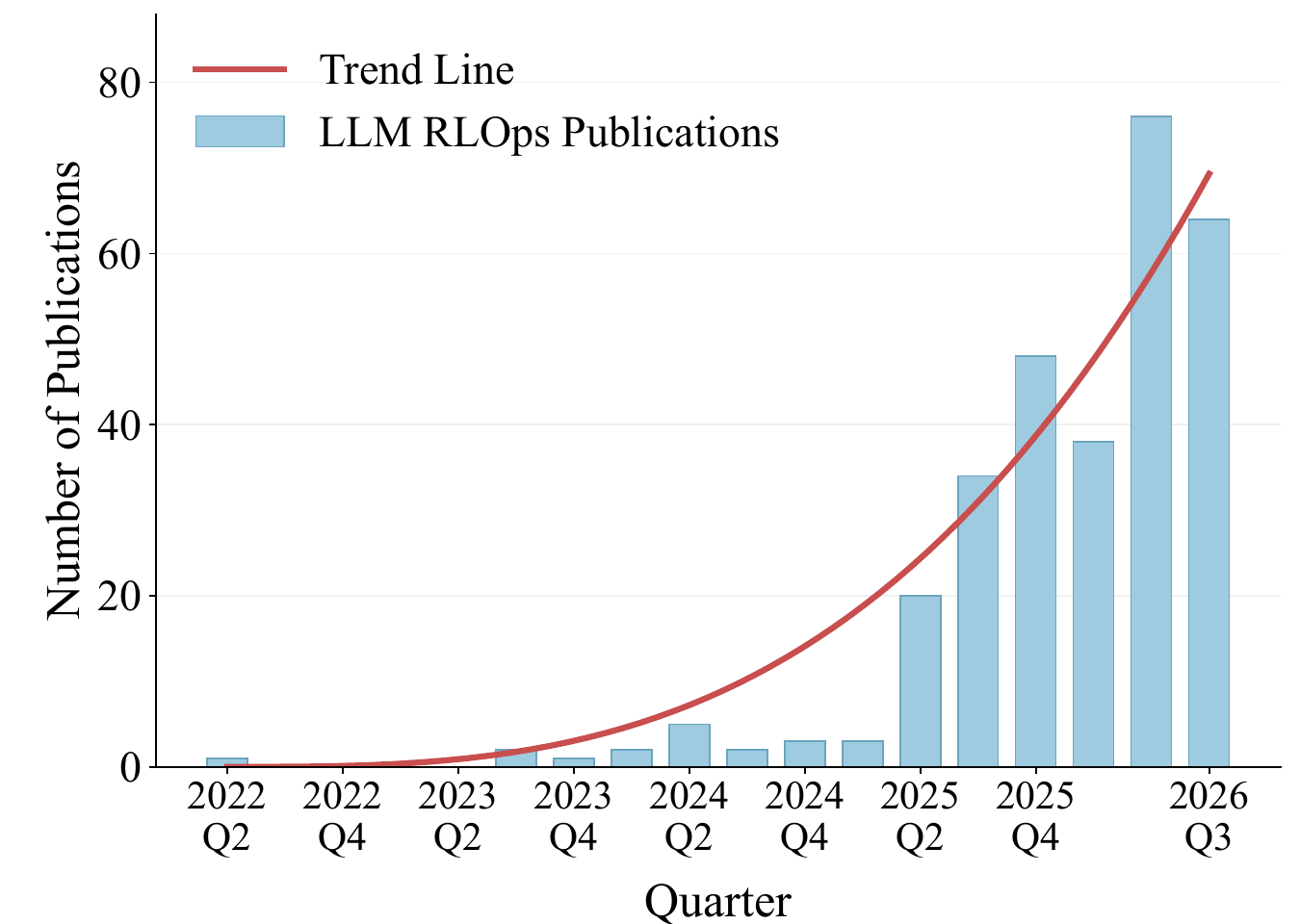}
			\label{fig: llm-ratio}
		\end{minipage}
	}
	\hspace{0.01\linewidth}
	\subfigure[Cumulative LLM and LLM RLOps publications.]{
		\begin{minipage}{0.47\linewidth}
			\centering
			\includegraphics[width=\textwidth]{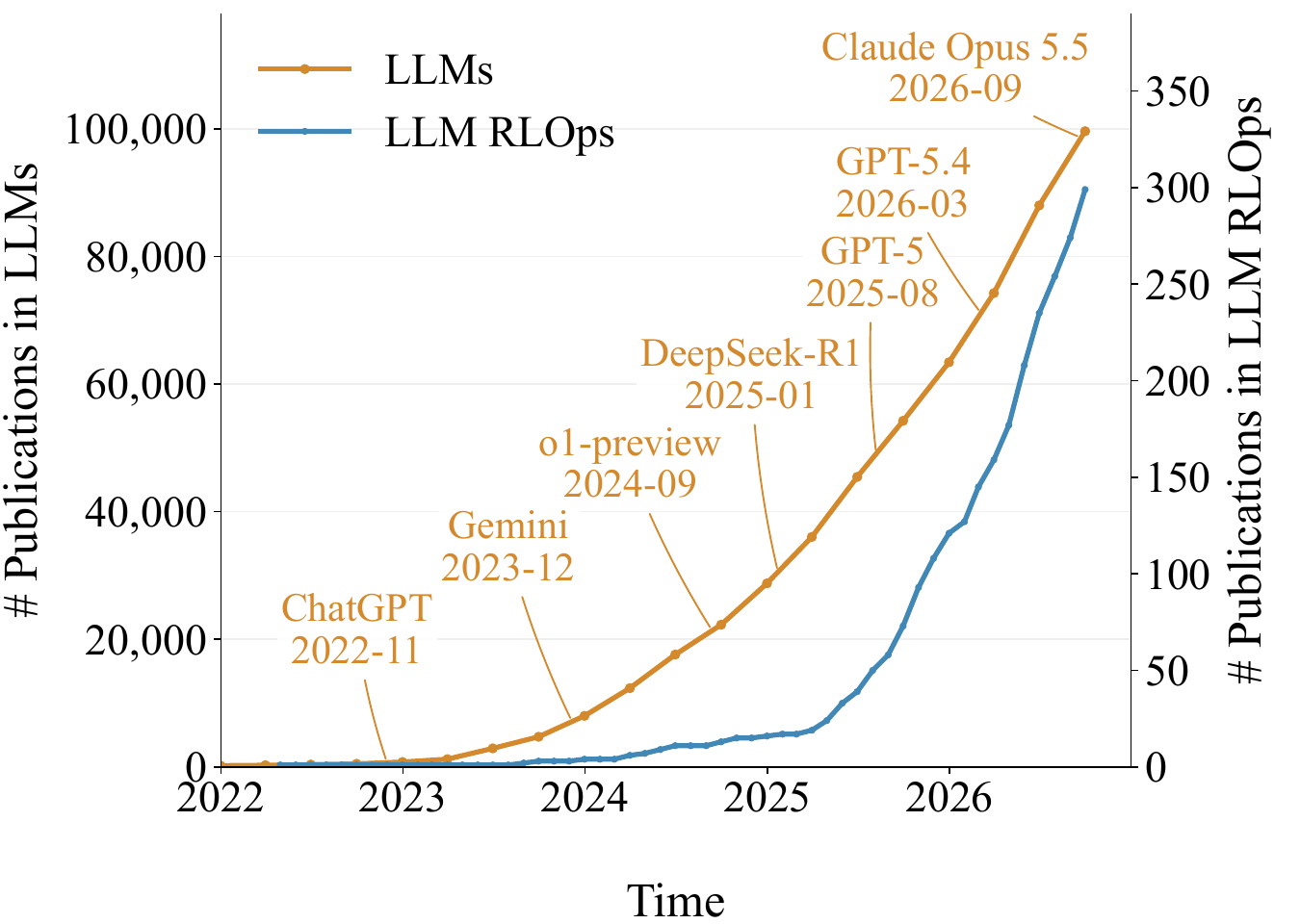}
			\label{fig: paper-num-arxiv}
		\end{minipage}
	}
	\caption{Analysis of Publication Trends in LLM RLOps}
	\label{fig: paper-num}
\end{figure}

The increasing adoption of RL in LLM development has stimulated growing research on managing its efficiency, stability, and reliability. As shown in Figure~\ref{fig: llm-ratio}, publications on LLM RLOps have increased substantially since 2025. This trend accompanies the broader expansion of LLM research, as illustrated in Figure~\ref{fig: paper-num-arxiv}. The growing literature motivates a systematic survey that consolidates existing approaches and clarifies the operational challenges of LLM reinforcement learning.

Existing reviews provide closely related perspectives on operations, reliability, and safety, as summarized in Table~\ref{tab: survey-comparison}. Early RLOps work examines the development, deployment, and monitoring of RL agents in Open RAN~\cite{li2022rlops}, while AgentOps organizes the monitoring, detection, localization, and resolution of anomalies in agent systems~\cite{wang2026agent}. These reviews establish important operational principles, but their primary focus differs from managing the coupled experience collection, reward computation, and policy updates of LLM reinforcement learning.

Recent reviews further examine the reliability and safety of self-evolving agents and recursive self-improvement (RSI). Wang et al.~\cite{wang2026diving} organize reliable self-evolution by evolution depth and emphasize independent evaluation and update acceptance controls. Gong et al.~\cite{gong2026evolutionary} examine safety preservation and risk propagation across successive updates, while Pu et al.~\cite{pu2026trustworthy} address improvement validity, constraint preservation, and effective oversight in RSI. Chen et al.~\cite{chen2026safety} investigate safety violations, containment, and repair as agents modify persistent state. These perspectives overlap with LLM RLOps in feedback reliability, update validation, and recovery, but cover broader forms of self-modification beyond RL training.

\begin{table}[htbp]
	\centering
	\caption{Comparison of closely related reviews.}
	\label{tab: survey-comparison}
	\begin{tabular}{ccp{10cm}<{\raggedright\arraybackslash}c}
		\toprule
		Reference & Year & Main Scope \\
		\midrule
		Li et al.~\cite{li2022rlops} & 2022 & RLOps: Lifecycle management, deployment, monitoring, and safety in Open RAN \\
		Wang et al.~\cite{wang2026agent} & 2026 & AgentOps: Monitoring, anomaly detection, root cause localization, and resolution in agent systems \\
		Wang et al.~\cite{wang2026diving} & 2026 & Reliable self-evolution: Evolution depth, independent evaluation, and update acceptance controls \\
		Gong et al.~\cite{gong2026evolutionary} & 2026 & Evolutionary safety: Safety preservation and risk propagation across updates and successor systems \\
		Pu et al.~\cite{pu2026trustworthy} & 2026 & Trustworthy RSI: Improvement validity, constraint preservation, and effective oversight \\
		Chen et al.~\cite{chen2026safety} & 2026 & Self-evolving agent safety: Persistent state changes, safety violations, containment, and repair \\
		\midrule
		\textbf{Our work} & --- & LLM RLOps: Operations management throughout the LLM reinforcement learning training lifecycle, including resource management, execution orchestration, monitoring and diagnosis, reliability control, failure recovery, and outcome validation \\
		\bottomrule
	\end{tabular}
\end{table}

Complementing these reviews, we focus specifically on operations management throughout the LLM reinforcement learning training lifecycle, including resource management, execution orchestration, monitoring and diagnosis, reliability control, failure recovery, and outcome validation. Our scope includes both conventional RL training and RL-based self-improvement, without requiring the training process itself to be recursively modified. We organize the literature into five stages: \textbf{Training Setup}, \textbf{Experience Collection}, \textbf{Reward Computation}, \textbf{Policy Optimization}, and \textbf{Model Evaluation}. Across these stages, we identify 18 operational topics and examine how system execution, feedback quality, and learning dynamics interact, providing a structured basis for comparing existing techniques and identifying open challenges.

\subsection{Structure of the Paper}

The remainder of this survey is organized as follows. Section~\ref{sec:systematic-review-process} describes our systematic review process. Section~\ref{sec:overview} introduces the necessary background and presents our taxonomy of LLM RLOps. Sections~\ref{sec:training-setup}--\ref{sec:model-evaluation} review operational challenges and techniques across training setup, experience collection, reward computation, policy optimization, and model evaluation. Section~\ref{sec:challenges} discusses open challenges and future research directions. Finally, Section~\ref{sec:conclusion} concludes the survey.

\section{Overview}
\label{sec:overview}

In this section, we introduce the background of reinforcement learning for LLM post-training and define the scope of LLM RLOps. We then present a taxonomy that organizes operational tasks across five stages of the training lifecycle.

\subsection{Background}
\label{sec:background}

\subsubsection{Reinforcement Learning for LLM Post-Training}
\label{sec:llm-rl-post-training}

Reinforcement learning (RL) adapts a pretrained LLM by using feedback to improve its generated responses or actions. The model acts as a policy that generates tokens conditioned on a prompt and the preceding interaction history. A reward function evaluates the resulting outputs, and a policy optimization algorithm updates the model to increase expected reward. Feedback can come from human preferences, AI evaluators, or verifiable outcomes such as executable tests.

The training process repeatedly collects experience, computes rewards, and updates the policy. Rollout workers generate responses or interact with environments to produce trajectories. These trajectories are scored and passed to the learner, which updates the model parameters and makes the resulting policy available for subsequent rollouts. Periodic evaluation assesses the resulting checkpoints and informs further training decisions.

This process couples generation, evaluation, and gradient computation within a distributed training loop. The components have different resource requirements and execution times: rollout duration varies with response length and environment interactions, reward computation may depend on additional models or external services, and policy updates require substantial memory and communication. Overlapping these operations can improve resource utilization, but also introduces delays between the policy used to collect experience and the policy being updated. Managing this loop therefore requires coordinating execution efficiency with experience freshness, feedback reliability, and learning stability.

\subsubsection{LLM RLOps}
\label{sec:llm-rlops}

RLOps extends MLOps practices to the development and operation of reinforcement learning systems. Earlier work, exemplified by RLOps for Open RAN, organizes activities such as model development, testing, deployment, monitoring, and safety management into a reproducible lifecycle~\cite{li2022rlops}. In this setting, operational concerns are organized around developing an RL agent and maintaining its behavior in an application environment.

In this survey, we define \emph{LLM RLOps} as the operational management of reinforcement learning runs for LLM post-training. Our unit of management is the training run and its interacting components: rollout workers, environments, reward evaluators, learners, and model checkpoints. Our scope covers the decisions and mechanisms needed to configure these components, coordinate their execution, maintain reliable learning signals, recover training progress, and validate the resulting model versions. Deployment and serving are considered when they directly support this training process.

This focus introduces a distinct operational setting. LLM post-training combines autoregressive generation with memory-intensive model updates, often across separate inference and training engines. Reward computation may itself require large models, executable verifiers, or interaction services. Meanwhile, successive policy updates change the trajectories collected and the feedback encountered, so system configuration and learning behavior continually influence each other. Efficient execution alone cannot establish that a run is producing useful model improvements.

Accordingly, our treatment of LLM RLOps centers on managing the coupled training loop. It connects resource and execution decisions with feedback quality, policy consistency, learning stability, and checkpoint acceptance. The following taxonomy organizes these responsibilities into Training Setup, Experience Collection, Reward Computation, Policy Optimization, and Model Evaluation.

\subsection{Taxonomy}
\label{sec:taxonomy}

\begin{figure}[htbp]
	\centering
	\begingroup
	\definecolor{stageP}{HTML}{8064A2}
	\definecolor{stageR}{HTML}{2980B9}
	\definecolor{stageW}{HTML}{C58A20}
	\definecolor{stageU}{HTML}{388657}
	\definecolor{stageV}{HTML}{C06060}
	\resizebox{\linewidth}{!}{%
		\begin{forest}
			for tree={draw,rounded corners=3pt,thin,/tikz/align=center,font=\scriptsize,anchor=west,parent anchor=east,child anchor=west,grow'=east,inner sep=3pt,l sep=12pt,s sep=3pt,fit=tight,edge={ultra thin},if n children=0{text width=5.3cm}{font=\bfseries\scriptsize,text width=3.0cm},edge path={\noexpand\path[draw,\forestoption{edge}] (!u.parent anchor) -- +(5pt,0) |- (.child anchor)\forestoption{edge label};}}
			[LLM RLOps,draw=gray,fill=gray!15,text width=1.7cm
			[Training Setup\\(\cref{sec:training-setup}),draw=stageP,fill=stageP!15
			[Computational Resources (\cref{sec:computation-resources}),draw=stageP,fill=stageP!6]
			[Run Configuration (\cref{sec:run-configuration}),draw=stageP,fill=stageP!6]
			[Base Model (\cref{sec:base-model}),draw=stageP,fill=stageP!6]
			[Training Data (\cref{sec:training-data}),draw=stageP,fill=stageP!6]
			]
			[Experience Collection\\(\cref{sec:experience-collection}),draw=stageR,fill=stageR!15
			[Sampling Budget Control (\cref{sec:sampling-budget-control}),draw=stageR,fill=stageR!6]
			[Rollout Execution Optimization (\cref{sec:rollout-execution-optimization}),draw=stageR,fill=stageR!6]
			[Environment Reliability (\cref{sec:environment-reliability}),draw=stageR,fill=stageR!6]
			[Trajectory Lifecycle Management (\cref{sec:trajectory-lifecycle-management}),draw=stageR,fill=stageR!6]
			]
			[Reward Computation\\(\cref{sec:reward-computation}),draw=stageW,fill=stageW!15
			[Learned Evaluators (\cref{sec:learned-evaluators}),draw=stageW,fill=stageW!6]
			[Programmatic Verifiers (\cref{sec:programmatic-verifiers}),draw=stageW,fill=stageW!6]
			[Reward Signals (\cref{sec:reward-signals}),draw=stageW,fill=stageW!6]
			]
			[Policy Optimization\\(\cref{sec:policy-optimization}),draw=stageU,fill=stageU!15
			[Update Execution (\cref{sec:update-execution}),draw=stageU,fill=stageU!6]
			[Policy Consistency (\cref{sec:policy-consistency}),draw=stageU,fill=stageU!6]
			[Learning Stability (\cref{sec:learning-stability}),draw=stageU,fill=stageU!6]
			[Training Recovery (\cref{sec:training-recovery}),draw=stageU,fill=stageU!6]
			]
			[Model Evaluation\\(\cref{sec:model-evaluation}),draw=stageV,fill=stageV!15
			[Capability Validation (\cref{sec:capability-validation}),draw=stageV,fill=stageV!6]
			[Risk Auditing (\cref{sec:risk-auditing}),draw=stageV,fill=stageV!6]
			[Version Acceptance (\cref{sec:version-acceptance}),draw=stageV,fill=stageV!6]
			]
			]
		\end{forest}%
	}
	\endgroup
	\caption{Taxonomy of LLM RLOps.}
	\label{fig:overview-taxonomy}
\end{figure}
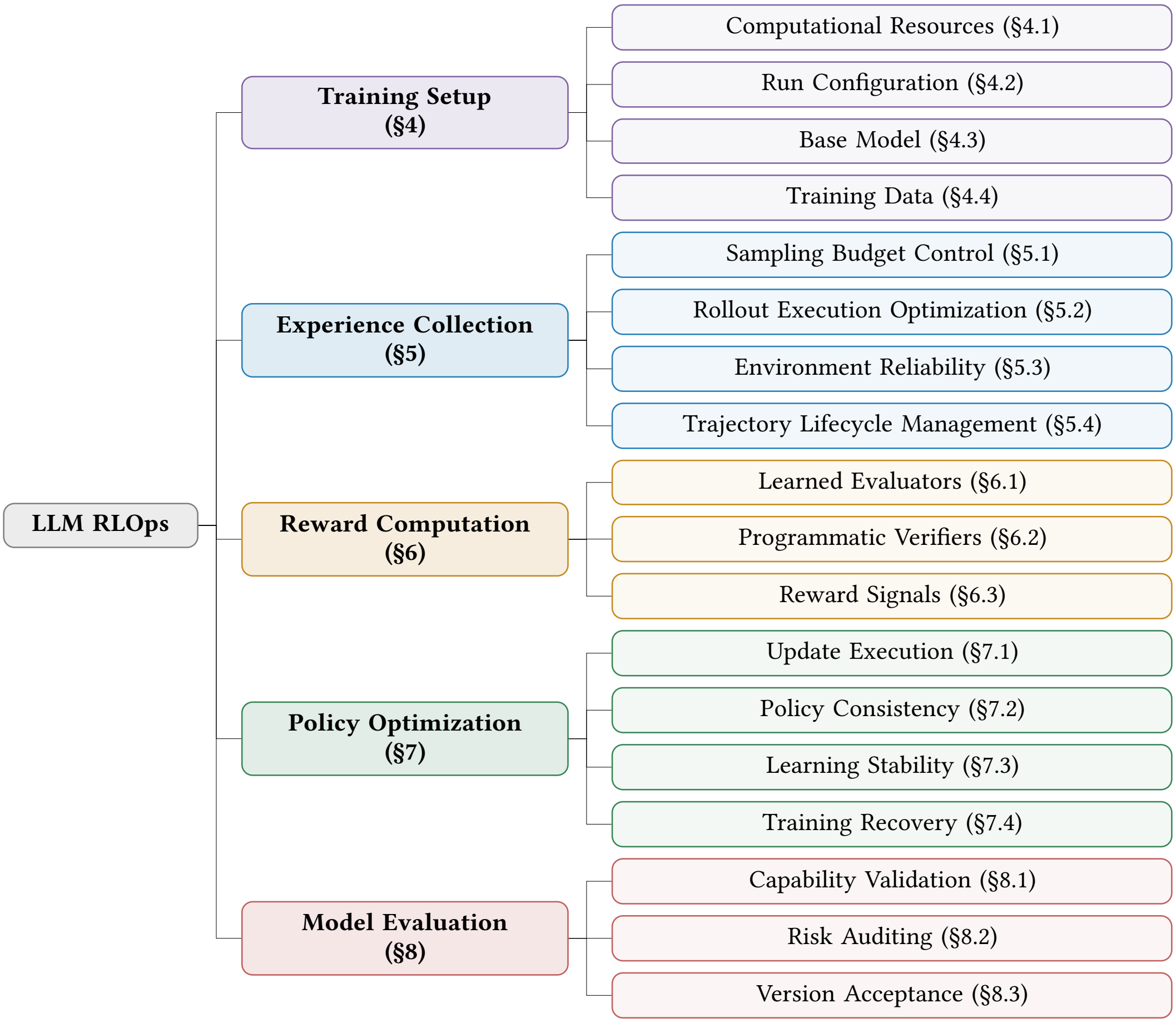

Building on the definitions above, we organize LLM RLOps research by the training stage at which its primary management contribution occurs. Figure~\ref{fig:overview-taxonomy} presents five stages and 18 operational topics. Within each stage, these topics distinguish the principal components and operational responsibilities involved in managing the training process.

Our taxonomy follows the LLM reinforcement learning lifecycle. \textbf{Training Setup} establishes the resources, configuration, initial model, and data for a training run. \textbf{Experience Collection} supplies trajectories, and \textbf{Reward Computation} evaluates them to produce training signals. \textbf{Policy Optimization} uses these signals to update the model, which generates subsequent experience. \textbf{Model Evaluation} assesses the resulting checkpoints and informs decisions to continue training, revise the setup, or accept a model version. Experience collection, reward computation, and policy optimization form a recurring training loop, while setup defines its operating conditions and evaluation guides its continuation.

\textbf{Training Setup} organizes the preparation of four interdependent objects. \textbf{Computational Resources} concerns the capacity and budget available for training. \textbf{Run Configuration} specifies the training recipe, component settings, and execution arrangements. \textbf{Base Model} concerns the initial policy and its suitability for subsequent RL, while \textbf{Training Data} concerns the tasks and supervision supplied to training. These objects jointly determine whether a proposed run is feasible and likely to support useful learning. Model size and task demands shape resource requirements; available resources constrain configuration choices; and the model's initial capabilities influence which data and training settings are appropriate.

\textbf{Experience Collection} coordinates the production and supply of training trajectories. \textbf{Sampling Budget Control} determines which tasks to sample, how much experience to generate, and when to stop generation. \textbf{Rollout Execution Optimization} executes the requested generation workload efficiently, while \textbf{Environment Reliability} maintains the external services and interaction state required by that workload. \textbf{Trajectory Lifecycle Management} governs the resulting records, including their admission, delivery, storage, reuse, and retirement. Together, these topics connect sampling demand to usable training experience: budget control sets the workload, rollout and environment operations support its execution, and lifecycle management controls the resulting supply to the learner.

\textbf{Reward Computation} connects feedback sources to the rewards used for learning. \textbf{Learned Evaluators} manages model-based scoring through calibration, uncertainty management, and robustness measures. \textbf{Programmatic Verifiers} manages executable checks through execution reliability, specification adequacy, and verifier integrity. These two sources can operate independently or jointly. \textbf{Reward Signals} manages how their outputs are composed across objectives, attributed to responses or individual steps, and corrected when necessary. The categories therefore distinguish the reliability of the scoring mechanisms from the construction of the resulting training rewards. Reliable feedback sources are necessary, but their outputs must also be combined and assigned appropriately to guide policy optimization.

\textbf{Policy Optimization} coordinates four operational responsibilities around model updates. \textbf{Update Execution} manages the placement, parallelism, memory, and state transfer required to execute updates efficiently. These execution choices can introduce version delays or numerical differences between rollout and training, which \textbf{Policy Consistency} measures and controls. \textbf{Learning Stability} addresses the resulting optimization dynamics, including gradient degradation, entropy collapse, and loss of useful learning signals. \textbf{Training Recovery} preserves or restores training state and progress when execution fails. Together, these topics address how updates are executed, whether collected experience remains compatible with learning, whether updates sustain useful progress, and how that progress survives interruptions.

\begin{table}[h]
	\centering
	\caption{RLOps methods coverage of representative LLM RL training frameworks. Numbers refer to the topics listed below.}
	\label{tab:framework-rlops-coverage}
	\begingroup
	\definecolor{opsSetup}{HTML}{8064A2}
	\definecolor{opsExperience}{HTML}{2980B9}
	\definecolor{opsReward}{HTML}{C58A20}
	\definecolor{opsPolicy}{HTML}{388657}
	\definecolor{opsEvaluation}{HTML}{C06060}
	\setlength{\tabcolsep}{3pt}
	\renewcommand{\arraystretch}{1.18}
	\footnotesize
	\begin{tabularx}{\linewidth}{@{}l*{5}{>{\centering\arraybackslash}X}@{}}
		\toprule
		\textbf{Framework}
		& \textcolor{opsSetup}{\makecell{Training\\Setup}}
		& \textcolor{opsExperience}{\makecell{Experience\\Collection}}
		& \textcolor{opsReward}{\makecell{Reward\\Computation}}
		& \textcolor{opsPolicy}{\makecell{Policy\\Optimization}}
		& \textcolor{opsEvaluation}{\makecell{Model\\Evaluation}} \\
		\midrule
		TRL~\cite{vonwerra2020trl} & 1, 2 & 2--4 & 2, 3 & 1--4 & 1, 3 \\
		trlX~\cite{havrilla2023trlx} & 2 & 2, 4 & \textemdash & 1--4 & 1, 3 \\
		DeepSpeed-Chat~\cite{yao2023deepspeed} & 2 & 2, 4 & \textemdash & 1--3 & \textemdash \\
		NeMo-Aligner~\cite{shen2024nemo} & 2 & 2 & \textemdash & 1--4 & 1, 3 \\
		NeMo RL~\cite{nemo-rl} & 2 & 1--4 & 2, 3 & 1--4 & 1, 3 \\
		verl~\cite{sheng2025hybridflow} & 2 & 1--4 & 2, 3 & 1--4 & 1 \\
		OpenRLHF~\cite{hu2024openrlhf} & 2 & 1--4 & 3 & 1--4 & 1, 3 \\
		AReaL~\cite{fu2026areal} & 2 & 1--4 & 2, 3 & 1--4 & 1 \\
		ROLL~\cite{lu2025rollflash} & 2 & 1--4 & 2, 3 & 1--4 & 1 \\
		Trinity-RFT~\cite{pan2025trinity} & 1, 2, 4 & 1--4 & 2, 3 & 1--4 & 1 \\
		RLinf~\cite{yu2025rlinf} & 1, 2 & 1--4 & 2, 3 & 1--4 & 1 \\
		slime~\cite{slime_github} & 2 & 1--4 & 3 & 1--4 & 1 \\
		SkyRL~\cite{griggs2025skrylv01} & 2 & 1--4 & 2, 3 & 1--4 & 1 \\
		AgentRL~\cite{zhang2025agentrl} & \textemdash & 2--4 & \textemdash & 1--4 & 1 \\
		
		\bottomrule
	\end{tabularx}
	\par\vspace{5pt}
	\fontsize{6.8}{8.2}\selectfont
	\renewcommand{\arraystretch}{1.05}
	\begin{tabularx}{\linewidth}{@{}p{0.20\linewidth}>{\raggedright\arraybackslash}X@{}}
		\textcolor{opsSetup}{Training Setup} & \textbf{1}~\hyperref[sec:computation-resources]{Computational Resources}; \textbf{2}~\hyperref[sec:run-configuration]{Run Configuration}; \textbf{3}~\hyperref[sec:base-model]{Base Model}; \textbf{4}~\hyperref[sec:training-data]{Training Data}. \\
		\textcolor{opsExperience}{Experience Collection} & \textbf{1}~\hyperref[sec:sampling-budget-control]{Sampling Budget Control}; \textbf{2}~\hyperref[sec:rollout-execution-optimization]{Rollout Execution Optimization}; \textbf{3}~\hyperref[sec:environment-reliability]{Environment Reliability}; \textbf{4}~\hyperref[sec:trajectory-lifecycle-management]{Trajectory Lifecycle Management}. \\
		\textcolor{opsReward}{Reward Computation} & \textbf{1}~\hyperref[sec:learned-evaluators]{Learned Evaluators}; \textbf{2}~\hyperref[sec:programmatic-verifiers]{Programmatic Verifiers}; \textbf{3}~\hyperref[sec:reward-signals]{Reward Signals}. \\
		\textcolor{opsPolicy}{Policy Optimization} & \textbf{1}~\hyperref[sec:update-execution]{Update Execution}; \textbf{2}~\hyperref[sec:policy-consistency]{Policy Consistency}; \textbf{3}~\hyperref[sec:learning-stability]{Learning Stability}; \textbf{4}~\hyperref[sec:training-recovery]{Training Recovery}. \\
		\textcolor{opsEvaluation}{Model Evaluation} & \textbf{1}~\hyperref[sec:capability-validation]{Capability Validation}; \textbf{2}~\hyperref[sec:risk-auditing]{Risk Auditing}; \textbf{3}~\hyperref[sec:version-acceptance]{Version Acceptance}. \\
	\end{tabularx}
	\endgroup
\end{table}

\textbf{Model Evaluation} connects evidence about trained models to version decisions. \textbf{Capability Validation} examines whether measured gains are valid, attributable to the changes under evaluation, and reproducible. \textbf{Risk Auditing} examines adverse changes that capability scores may conceal, including behavioral regressions, unsafe behavior, data exposure, and limitations in reasoning monitorability. These two categories provide complementary evidence about benefits and risks under comparable evaluation conditions. \textbf{Version Acceptance} uses this evidence to select checkpoints and determine whether model or configuration changes should be retained. Accepted versions provide references for subsequent comparisons, while rejected changes can prompt further training or revisions to earlier stages.

To connect this taxonomy with existing implementations, Table~\ref{tab:framework-rlops-coverage} summarizes the operational topics supported by representative LLM RL training frameworks. Each entry identifies a topic for which the reviewed implementation provides a concrete management mechanism, including mechanisms available through integrated backends or shipped training recipes. The comparison provides an implementation-oriented reference for the techniques reviewed in the following sections.

The reviewed frameworks commonly support rollout execution, distributed policy updates, policy consistency, and training-state recovery. Support for resource planning and evaluation-based version selection is established in a smaller subset of the reviewed implementations. These differences reflect the operational scope of the frameworks, while topic-level coverage does not establish equivalent implementation depth or effectiveness. The following sections examine the mechanisms within each topic, including specialized approaches that may not yet be integrated into general-purpose training frameworks.

\section{Systematic Review Process}
\label{sec:systematic-review-process}

We establish a review protocol covering the search strategy, inclusion and exclusion criteria, and publication selection process~\cite{khan2003five, kitchenham2009systematic}. This protocol guides a transparent and reproducible review.

\subsection{Search Strategy}
\label{sec:search-strategy}

We search four sources: the ACM Digital Library, IEEE Xplore, Web of Science, and arXiv\footnote{\url{https://dl.acm.org/}, \url{https://ieeexplore.ieee.org/}, \url{https://www.webofscience.com/}, and \url{https://arxiv.org/}.}. These sources cover software engineering, computer systems, and machine learning, including both published studies and preprints. Records are collected through multiple search rounds.

\begin{figure}[h]
	\centering
	\includegraphics[width=\textwidth]{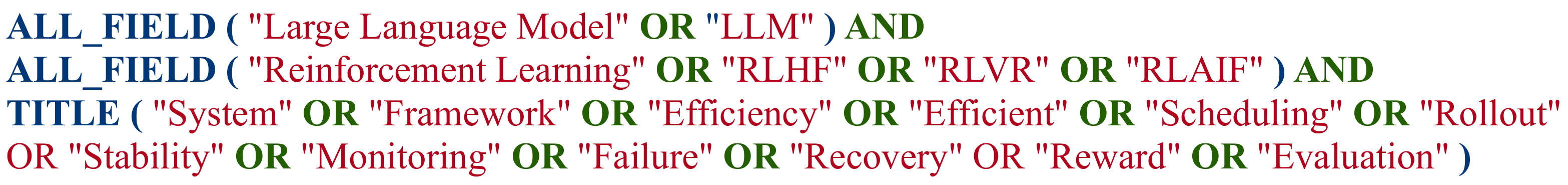}
	\caption{Search Strategy Utilized to Identify Studies on RLOps}
	\label{fig:search}
\end{figure}

Based on a preliminary investigation, we formulate the search string shown in Figure~\ref{fig:search}. It combines two groups of terms: LLM reinforcement learning and training operations. The first includes RLHF, RLVR, and RLAIF, as well as combinations of language-model and reinforcement-learning terms. The second covers training infrastructure, resource efficiency, stability, monitoring, recovery, and feedback reliability. We do not require the term \emph{RLOps}, as relevant studies may address these concerns without using this label. Retrieved records are subsequently screened using the inclusion and exclusion criteria.

\subsection{Selection Criteria}
\label{sec:selection-criteria}

Papers are selected based on their relevance to the research questions and scope of this survey. We focus on the operation and management of LLM reinforcement learning, including its supporting infrastructure. Papers must satisfy at least one of the following inclusion criteria:

\begin{itemize}
	\item \textbf{IC1:} The paper improves the execution or resource efficiency of LLM reinforcement learning.
	\item \textbf{IC2:} The paper manages training data, experience, or interaction environments.
	\item \textbf{IC3:} The paper identifies or mitigates failures in training feedback or policy updates.
	\item \textbf{IC4:} The paper supports training monitoring, evaluation, recovery, or iteration.
\end{itemize}

We initially use an LLM to screen titles and abstracts against the selection criteria. These preliminary decisions guide subsequent reviewer assessment rather than determine final inclusion. Full texts are examined to confirm eligibility and resolve uncertain cases. Papers are excluded if they meet any of the following conditions:

\begin{itemize}
	\item \textbf{EC1:} The study addresses neither LLM reinforcement learning nor its supporting infrastructure.
	\item \textbf{EC2:} The study lacks a direct contribution to managing LLM reinforcement learning.
	\item \textbf{EC3:} The study improves task performance without addressing training operations.
	\item \textbf{EC4:} The item is not a standalone research contribution.
\end{itemize}

Each record is reviewed by a primary reviewer, with non-trivial cases discussed and resolved with at least one secondary reviewer. The final set of selected papers is validated by all co-authors. The criteria distinguish training operations from ordinary task improvement. Studies must address an operational concern in LLM reinforcement learning, rather than merely use RL to improve downstream performance. Supporting infrastructure is considered when it directly serves the training process. Studies of unrelated systems or inference-only workloads are excluded.

\begin{figure}[h]
	\centering
	\includegraphics[width=\textwidth]{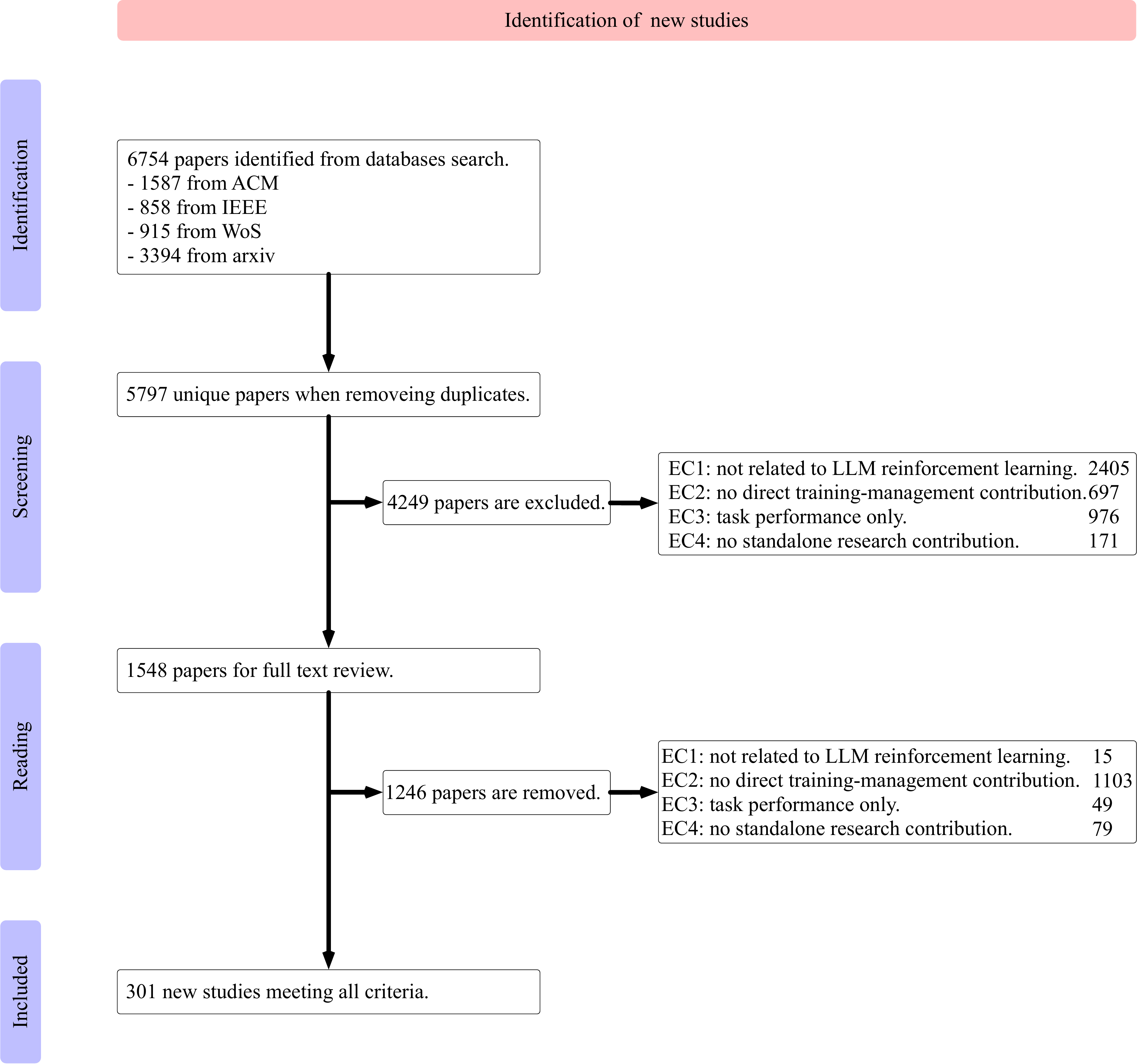}
	\caption{Overview of Paper Selection Procedure}
	\label{fig:selected-publications}
\end{figure}

\subsection{Overview of Selected Publications}
\label{sec:overview-selected-publications}

Figure~\ref{fig:selected-publications} summarizes the construction of the current corpus. Across all recorded retrieval rounds, we collect 6,754 source records: 1,587 from ACM, 858 from IEEE Xplore, 915 from Web of Science, and 3,394 from arXiv. Exact deduplication removes 957 records, leaving 5,797. 

During title and abstract screening, we remove 4,249 records from the candidate pool based on criteria EC1–EC4 or their background or unresolved status. We then seek full texts for the remaining 1,548 candidates and obtain usable text for 1,491. Of these, 1,170 are excluded, primarily because they lack a direct contribution to training management. The resulting 301 studies form the core corpus for this review

\section{Training Setup}
\label{sec:training-setup}

Training setup establishes the resources, artifacts, and configuration required for an LLM reinforcement learning run. Decisions made at this stage affect both execution feasibility and subsequent learning: insufficient resources restrict the training scale, unsuitable configurations waste computation, and poorly matched models and data can limit useful learning signals. The operational objective is therefore to assemble a feasible setup with a justified resource budget and suitable conditions for improvement.

\begin{figure}[htbp]
	\centering
	\includegraphics[width=\textwidth]{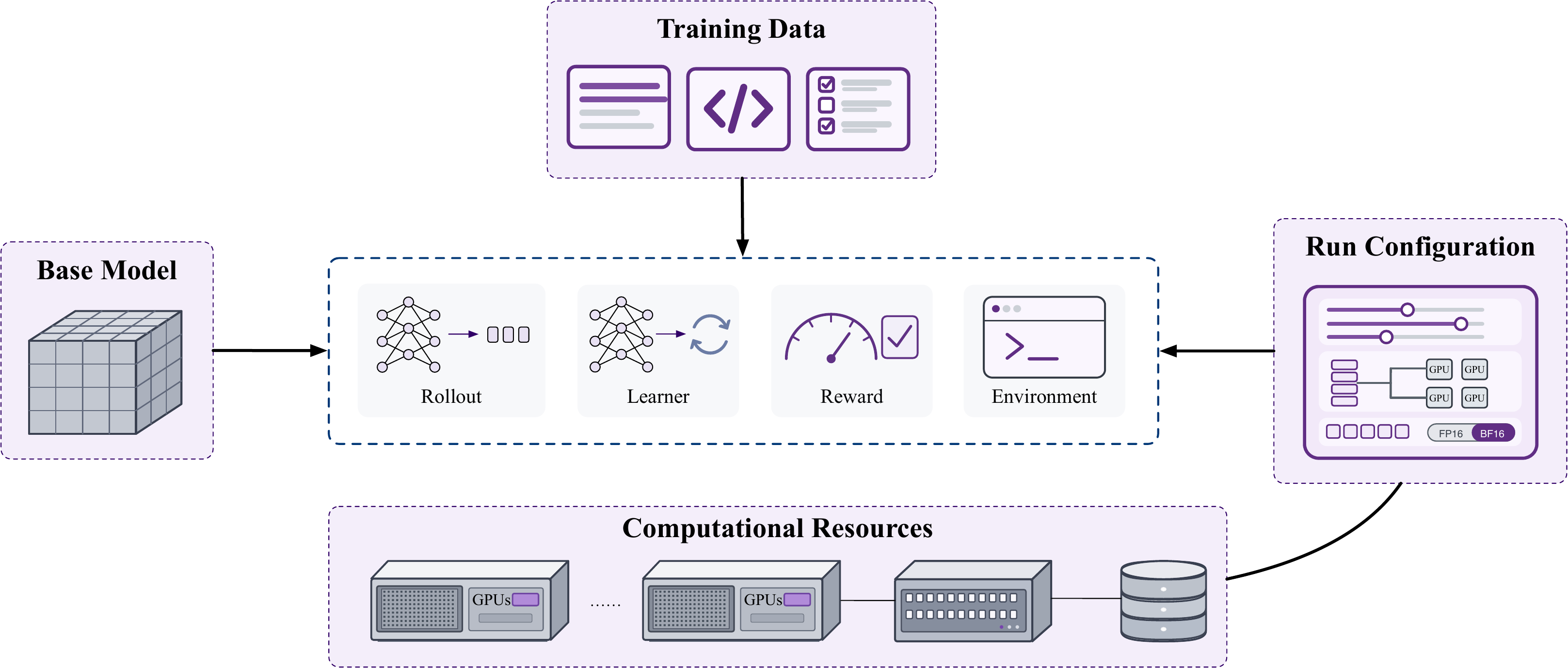}
	\caption{Overview of Training Setup in LLM RLOps. Computational resources host the training components, the base model initializes the policy, and training data provides task inputs and scoring references. Run configuration governs component behavior and resource allocation.}
	\label{fig:training-setup}
\end{figure}

As illustrated in Figure~\ref{fig:training-setup}, we organize this stage around four objects: \textbf{Computational Resources}, \textbf{Run Configuration}, \textbf{Base Model}, and \textbf{Training Data}. Computational resources provide the capacity to execute the run, while run configuration specifies how training components use that capacity. The base model supplies the initial policy, and training data defines the tasks and associated supervision available to it. These objects constrain one another: model size and task characteristics affect resource requirements, available hardware constrains configuration choices, and the initial policy's capabilities determine which training tasks provide useful learning opportunities.

\subsection{Computational Resources}
\label{sec:computation-resources}

Computational resources constrain the scale and cost of LLM reinforcement learning. Existing work addresses three aspects of resource planning: estimating workload requirements, predicting the returns from additional computation, and allocating limited training budgets.

\paragraph{Resource Demand Estimation.}

Resource estimation accounts for generation, reward computation, policy updates, and environment interactions. Profiling model-role bottlenecks and memory use informs resource assignments~\cite{xiao2023adaptive}, while coding-agent measurements expose additional environment initialization and execution costs~\cite{graviet2026rollout}. ECHO-2~\cite{song2026echo2} connects these estimates to rollout provisioning: its capacity rule combines learner step time, weight-dissemination latency, worker throughput, and an allowed staleness budget to estimate the rollout capacity needed to keep the learner supplied. Resource requirements therefore depend on both workload cost and the coordination constraints imposed on execution.

\paragraph{Compute--Performance Prediction.}
Several studies characterize how training performance changes with computational investment, model scale, and data volume~\cite{khatri2026art,nimmaturi2025predictive,tan2026scaling,mai2025agent}. For example, ScaleRL~\cite{khatri2026art} fits compute--performance curves to extrapolate learning gains from smaller-scale experiments, while Nimmaturi et al.~\cite{nimmaturi2025predictive} model reward saturation to identify diminishing returns from continued GRPO training. These approaches support decisions about training scale and duration.

\paragraph{Compute Budget Allocation.}
Other studies examine how to distribute computation across model scale, response sampling, and training iterations~\cite{hou2024does,cheng2026isocompute}. Hou et al.~\cite{hou2024does} compare investments in RLHF components, revealing diminishing returns from additional sampling and larger reward models. Cheng et al.~\cite{cheng2026isocompute} study fixed-budget allocation across problems per batch, rollouts per problem, and update steps. Together, these studies support choosing an allocation according to the limiting factors of the training setup.

\subsection{Run Configuration}
\label{sec:run-configuration}

Run configuration specifies the learning algorithms, hyperparameters, component combinations, and execution settings of an RL run. Existing work investigates how to identify effective configurations through empirical analysis, automate their selection, and validate their execution prerequisites.

\paragraph{Empirical Configuration Analysis.}
Several studies use controlled comparisons and ablations to identify effective training settings and their interactions~\cite{ivison2024unpacking,vattikonda2026train,wang2025nemotron,he2025justrl,hao2026nebulaexp,li2026larger}. These analyses cover both complete training recipes and individual choices, such as batch size, learning rate, and regularization. For example, Ivison et al.~\cite{ivison2024unpacking} separate the effects of learning algorithms, preference data, reward models, and policy training prompts. He et al.~\cite{he2025justrl} examine whether additional training techniques improve a simpler recipe, while Li et al.~\cite{li2026larger} investigate the interaction of batch size with learning rates and execution throughput. Such evidence helps distinguish useful configuration choices from unnecessary or harmful combinations.

\paragraph{Automated Configuration Search.}
Another line of work reduces the cost of finding suitable configurations through automated search~\cite{chen2026efficient,chwa2026automatic,sheng2025hybridflow}. Chen et al.~\cite{chen2026efficient} combine Bayesian optimization with evaluations at different model scales and training budgets to avoid fully training every candidate. Chwa et al.~\cite{chwa2026automatic} further incorporate transferable configuration knowledge and early performance prediction. At the systems level, HybridFlow~\cite{sheng2025hybridflow} searches device mappings and parallelization strategies, extending configuration selection from learning parameters to execution arrangements.

\paragraph{Configuration Validation.}
Configuration validation checks whether the selected settings satisfy the prerequisites for correct execution. Relevant work addresses component compatibility, backend constraints, and parameter-update paths~\cite{lyngkhoi2026aligntune,robustagents2026}. AlignTune~\cite{lyngkhoi2026aligntune} provides configuration validation within a modular post-training toolkit, while work on robust RL for small language model agents identifies adapter and gradient-path failures that can prevent intended parameters from being updated~\cite{robustagents2026}. These checks help detect configurations that can launch a training process but fail to execute the intended learning procedure.

\subsection{Base Model}
\label{sec:base-model}

The base model establishes the starting capabilities and behavior distribution of an RL run. Its management involves selecting a suitable model, preparing it for reinforcement learning, and identifying a checkpoint that supports subsequent improvement. Existing studies show that these decisions require considering both current performance and the model's response to further training.

\paragraph{Model Selection.}
Several studies examine how model family, scale, and prior training affect RL outcomes~\cite{zeng2025simplerl,hou2024does,tan2026scaling}. Zeng et al.~\cite{zeng2025simplerl}, for example, compare zero-RL training across diverse model families and sizes, finding distinct training dynamics and dependencies on initial capabilities and task difficulty. Scaling studies further reveal interactions between policy size, reward models, and computational investment~\cite{hou2024does,tan2026scaling}. These findings support selecting the starting model jointly with the intended workload and feedback mechanism.

\paragraph{Initialization Preparation.}
Another line of work prepares the initial policy to acquire task-specific behavior while preserving its capacity for further learning~\cite{zhang2026good,liu2026clipping,liu2026rl,cheng2026teaching}. Zhang et al.~\cite{zhang2026good} propose PEAR, which reweights the SFT objective to account for the mismatch between offline demonstrations and subsequent on-policy learning. Other approaches constrain SFT-induced changes or repair their effects: token-level clipping preserves prior knowledge during adaptation~\cite{liu2026clipping}, while Rejuvenation combines base-anchored model fusion with targeted neuron resets to restore plasticity after excessive SFT~\cite{liu2026rl}. For tool-integrated reasoning, Cheng et al.~\cite{cheng2026teaching} study preparation that develops tool-use behavior while retaining reasoning capabilities.

\paragraph{Checkpoint Selection.}
Checkpoint selection identifies which available model state should enter RL, using indicators of subsequent learning potential~\cite{kang2026quagmires,aphale2026sft,cheng2026teaching}. Kang et al.~\cite{kang2026quagmires} show that higher SFT scores do not reliably predict better post-RL performance and identify held-out generalization loss and pass@k at large k as more informative predictors. Aphale et al.~\cite{aphale2026sft} further associate poor downstream GRPO performance with reduced output diversity in overtrained checkpoints, proposing pre-RL entropy screening alongside early training monitoring. These approaches make checkpoint selection sensitive to the exploration and generalization properties needed by the subsequent RL run.

\subsection{Training Data}
\label{sec:training-data}

Training data defines the tasks and supervision available to an RL run. Its preparation involves selecting informative examples, validating their quality, and allocating annotation resources. These decisions determine whether the training inputs provide useful and trustworthy learning signals for the selected model.

\paragraph{Informativeness-Based Data Selection.}
Several studies select training subsets according to task difficulty, model uncertainty, and expected learning value~\cite{zhao2026ufo,li2025truth,team2025intellect}. UFO-RL~\cite{zhao2026ufo}, for example, uses single-pass uncertainty estimation to identify informative examples without expensive repeated sampling. For multimodal reasoning, Li et al.~\cite{li2025truth} select examples that require meaningful use of visual information, filtering out those dominated by language priors or irrelevant attention. INTELLECT-2~\cite{team2025intellect} applies offline difficulty filtering to remove problems that are too easy or too difficult for the starting model. These approaches align the initial training collection with the model's learning capabilities.

\paragraph{Data Quality Validation.}

Data validation checks example quality and whether generated tasks can be executed and verified~\cite{jing2026guiding,zhou2026synthetic,chu2026redsearcher}. Jing et al.~\cite{jing2026guiding} use sparse-autoencoder representations to construct a quality probe, while agent-training pipelines apply sanity checks to synthesized tasks~\cite{zhou2026synthetic,chu2026redsearcher}. This requirement extends to inputs regenerated during training. In controlled self-play experiments, Pu et al.~\cite{pu2026survive} find that execution-grounded task admission prevents collapse across the reward variants tested, whereas changing rewards alone does not compensate for removing the admission gate. Their results distinguish validating which tasks enter the pool from scoring responses to admitted tasks.

\paragraph{Annotation Budget Allocation.}
A further line of work manages the cost of obtaining supervision, deciding which examples warrant annotation and how annotation budgets should be distributed~\cite{zhu2026smart,wang2026scaling}. Zhu et al.~\cite{zhu2026smart} propose PivotTrace, which uses reasoning uncertainty to route examples toward annotation, self-supervision, or exclusion. At the training-pipeline level, Wang et al.~\cite{wang2026scaling} study how to divide annotation budgets between SFT and RL, using small proxy models to identify allocation regions that transfer to larger models. These approaches connect supervision expenditure to expected training value.

\section{Experience Collection}
\label{sec:experience-collection}

Experience collection produces the trajectories used for reward computation and policy optimization. Rollout workers generate responses or interact with environments using a model checkpoint, recording the resulting tokens, actions, and observations. Managing this process requires balancing the value and cost of new experience, coordinating variable-length generation workloads, maintaining reliable interactions, and delivering usable trajectories to subsequent training stages.

\begin{figure}[htbp]
	\centering
	\includegraphics[width=\textwidth]{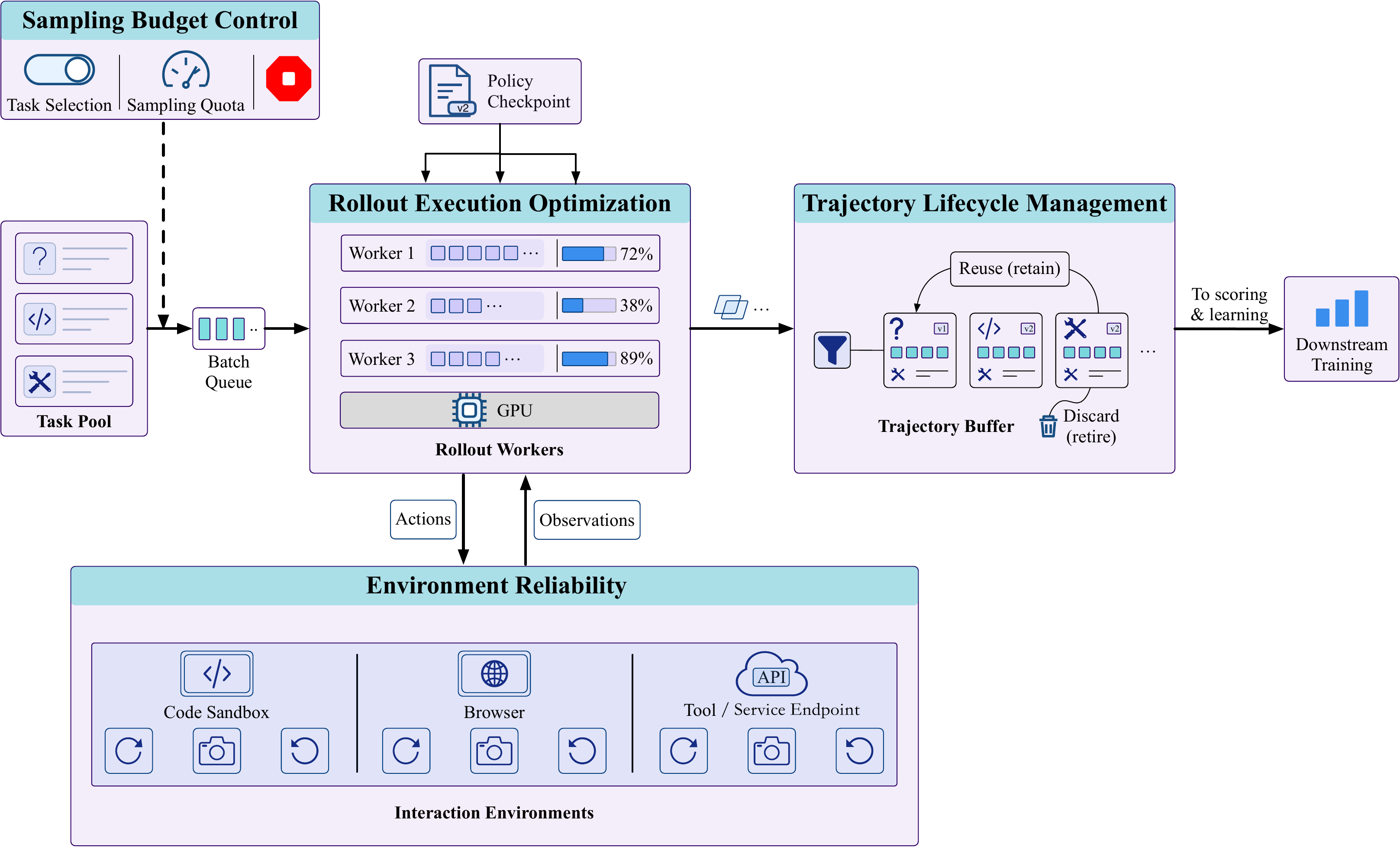}
	\caption{Overview of Experience Collection in LLM RLOps. Sampling control directs task dispatch to rollout workers, which generate trajectories through model inference and environment interaction. Trajectory management organizes collected experience for downstream training and provides feedback for subsequent sampling decisions.}
	\label{fig:experience-collection}
\end{figure}

As illustrated in Figure~\ref{fig:experience-collection}, we organize these responsibilities into four topics. \textbf{Sampling Budget Control} determines which tasks receive sampling effort, how many trajectories to generate, and when to stop. \textbf{Rollout Execution Optimization} manages the execution of these sampling requests through batching, concurrency, and inference-resource utilization. \textbf{Environment Reliability} maintains the services and state required for successful interactions. \textbf{Trajectory Lifecycle Management} governs the admission, storage, delivery, reuse, and retirement of collected experience. These responsibilities are connected through the collection process: sampling decisions shape the workload, worker and environment execution produce trajectories, and the availability of collected experience informs subsequent sampling needs.

\subsection{Sampling Budget Control}
\label{sec:sampling-budget-control}

Sampling budget control determines where and how much experience to collect as the policy evolves. Uniform sampling can repeatedly generate trajectories that provide little learning signal, particularly when all responses to a prompt receive identical rewards. Existing work controls this expenditure through online task selection, adaptive rollout allocation, and early termination.

\paragraph{Online Task Selection.}

Online selection prioritizes tasks using difficulty estimates, reward histories, or observed failures~\cite{shen2026bots,qu2026can,gao2026prompt,zhu2026kalman,wang2026sentinel,zheng2025act}. BOTS~\cite{shen2026bots} maintains Bayesian difficulty estimates and uses Thompson sampling to balance exploration and exploitation. GRESO~\cite{zheng2025act} instead uses reward histories to skip prompts likely to yield no within-group reward variation before generating their responses, while retaining occasional exploration as the policy changes.

Sampling must also control the task mixture actually delivered to training. Faster domains can dominate completed trajectories even when dispatch proportions match the intended mixture. Apriel-Reasoner~\cite{pardinas2026apriel} adjusts domain sampling using completed-rollout proportions, connecting selection decisions to execution feedback rather than relying on a fixed dispatch distribution.

\paragraph{Adaptive Rollout Allocation.}
Another line of work varies the number of trajectories generated for each task rather than assigning a fixed group size~\cite{zhang2026train,wang2026spend,kim2026spend,jiang2026learning,zou2026trace}. These approaches use preliminary samples or accumulated training evidence to estimate where additional generation is likely to be useful. Wang et al.~\cite{wang2026spend}, for example, allocate rollout budgets according to the estimated utility of obtaining successful responses, while progressive allocation methods increase sampling for groups that retain informative reward variation~\cite{jiang2026learning}. TRACE~\cite{zou2026trace} extends allocation to multi-turn interactions, distributing a fixed budget across both prompts and intermediate trajectory prefixes to improve reward contrast.

\paragraph{Early Termination and Pruning.}
Complementary approaches use evidence gathered during collection to decide when further sampling should stop~\cite{xu2026prune,zhai2026selective,voss2026early}. Xu et al.~\cite{xu2026prune} train a lightweight quality head to estimate the success probability of partial responses and prune rollouts during generation. Zhai and Wang~\cite{zhai2026selective} use similarity between partial action sequences to identify groups likely to produce identical rewards and terminate them early. Sequential allocation methods also decide whether to continue collecting responses or close a sampling group~\cite{voss2026early}. These interventions avoid completing low-value generation while seeking to preserve useful training signals.

\subsection{Rollout Execution Optimization}
\label{sec:rollout-execution-optimization}

Rollout execution optimization improves the efficiency of generating the requested experience. Variable response lengths, multi-turn interactions, and changing resource demands can leave accelerators idle or create execution bottlenecks. Existing work addresses these problems through workload scheduling, asynchronous execution, generation acceleration, and adaptive resource management.

\paragraph{Long-Tail Scheduling.}
Scheduling reduces waiting from uneven rollout lengths through workload grouping and continuous sampling~\cite{gao2025rollpacker,zhang2026sortedrl,wang2025infinite}. RollPacker~\cite{gao2025rollpacker} groups long responses into designated rounds, while SortedRL~\cite{zhang2026sortedrl} organizes completed samples for earlier updates. Partial-rollout approaches pause unfinished responses and resume them later~\cite{zhou2025april,kimi2025k15,qu2025copris}. APRIL~\cite{zhou2025april}, for example, over-provisions requests and proceeds once enough completed samples are available. These mechanisms shorten synchronization waits while retaining unfinished work.

\paragraph{Asynchronous and Streaming Execution.}

Asynchronous systems decouple rollout production from policy updates, allowing generation to continue while the learner consumes available batches~\cite{fu2026areal,hu2024openrlhf,wang2025distrl,lu2025rollflash,zhang2025agentrl,cao2025skyrlagent}. Streaming systems further overlap stages or process partial outputs~\cite{zhong2025streamrl,hu2026dora,yan2025oppo}. StreamRL~\cite{zhong2025streamrl} overlaps pipeline stages, while DORA~\cite{hu2026dora} maintains multiple policy versions. The resulting utilization gains require explicit control of experience staleness.

\paragraph{Generation Acceleration.}
Several approaches reduce the computation required to produce rollout tokens through speculative decoding and low-precision execution~\cite{liu2025specrl,he2025rhymerl,zhang2025fastgrpo,chen2025respec,huang2025qerl}. SPEC-RL and RhymeRL~\cite{liu2025specrl,he2025rhymerl} exploit similarities between successive training iterations, using historical responses to construct drafts that are verified by the current policy. FastGRPO~\cite{zhang2025fastgrpo} adapts drafting and verification to changing concurrency and updates the drafter as the policy evolves. QeRL~\cite{huang2025qerl} reduces rollout computation and memory costs through low-precision quantization, while also examining its effects on exploration.

\paragraph{Adaptive Resource Management.}
Other systems adapt rollout capacity and parallelism to changing workloads~\cite{tan2025earl,wu2025rlboost,gao2026rose,wei2026rlhfless,du2026dynaresize}. EARL~\cite{tan2025earl} adjusts parallelism to context length and load, while RLBoost~\cite{wu2025rlboost} uses preemptible GPUs with adaptive offloading and response migration. ROSE~\cite{gao2026rose} uses spare serving capacity while protecting serving objectives. Related approaches use cost-aware serverless scaling or move GPUs between rollout and training roles~\cite{wei2026rlhfless,du2026dynaresize}. Resource adaptation must account for transition overhead as well as the imbalance it seeks to remove.

\subsection{Environment Reliability}
\label{sec:environment-reliability}

Environment reliability maintains the execution services, interaction behavior, and state required for multi-turn rollouts. Unavailable tools, unrepresentative simulated responses, or lost state can disrupt collection or distort learning. Existing work addresses environment provisioning, interaction availability and fidelity, and state recovery.

\paragraph{Environment Provisioning and Isolation.}
Several systems manage the creation, allocation, and reuse of environments to support concurrent agent rollouts~\cite{huang2026dsec,wang2025agentfly,xiao2026tangram,zhang2026prorlagent}. DSec~\cite{huang2026dsec}, for example, provides multiple sandbox backends through a unified interface and coordinates their placement and lifecycle, using on-demand image loading to support large bursts of environment creation. AgentFly~\cite{wang2025agentfly} manages environment allocation and reuse through a centralized resource system. ARL-Tangram~\cite{xiao2026tangram} further allocates external resources at the action level, avoiding the need to reserve them throughout an entire trajectory. These mechanisms help maintain environment availability under changing interaction workloads.

\paragraph{Interaction Availability and Reproducibility.}

Controlled environments, interaction replay, and resilient tool interfaces reduce dependence on unstable external services~\cite{kar2026weblica,chu2026redsearcher,zhou2026synthetic,li2025websailor,chai2025rlfactory}. Weblica~\cite{kar2026weblica} uses HTTP-level caching and replay to reproduce web states while preserving interactive behavior. For live services, WebSailor-V2~\cite{li2025websailor} combines rate limits, caching, retries, and backup sources to reduce API disruptions.

\paragraph{Simulation Fidelity.}
An available simulator can still provide an increasingly unrepresentative training environment. Yu et al.~\cite{yu2026simulator} observe rising training rewards alongside declining held-out performance when policies repeatedly interact with one frozen simulator. They diversify simulated responses or co-train a population of simulators to reduce overfitting. This adds a behavioral reliability requirement: simulated interactions must retain sufficient diversity to support transfer beyond the training simulator.

\paragraph{State Checkpointing and Recovery.}
Stateful rollouts require environment recovery to match the agent's recorded context~\cite{dong2026deltabox,wu2026crab,zhou2026belayer}. DeltaBox~\cite{dong2026deltabox} accelerates sandbox checkpointing and rollback by tracking changes in filesystem and process state. Crab~\cite{wu2026crab} uses the operating-system effects of agent actions to determine when and what to checkpoint, avoiding unnecessary state capture. Belayer~\cite{zhou2026belayer} coordinates environment restoration with the corresponding LLM context so that interrupted rollouts resume from a consistent interaction prefix. Together, these mechanisms preserve completed interaction work and prevent recovery from exposing agents to mismatched environment states.

\subsection{Trajectory Lifecycle Management}
\label{sec:trajectory-lifecycle-management}

Trajectory lifecycle management governs how collected experience is represented, stored, delivered, reused, and retired. Generated trajectories must preserve the interaction information required for learning, remain accessible across distributed components, and be assessed for continued usefulness as the policy changes. Existing work addresses these requirements through trajectory representation and delivery, admission control, and experience replay.

\paragraph{Trajectory Representation and Delivery.}

Systems standardize interaction records and their transfer from agent runtimes to learners~\cite{luo2025lightning,xu2026polar,pan2025trinity,le2025toolbrain,cheng2025agentr1}. Agent Lightning v1.0~\cite{he2026lightning} shows why reconstruction affects learning: retokenization and sample splitting can turn one rollout into multiple training rows, so sample-level weighting can unintentionally favor rollouts with more rows. It coordinates rollout-level advantage calculation and loss normalization to preserve the intended weighting. Storage and transfer systems complement this representation layer with inspectable records, distributed storage, and streaming delivery~\cite{han2025asyncflow,wang2026clawr1,pan2025trinity}.

\paragraph{Admission and Freshness Control.}

Admission control selects collected trajectories according to learning value and freshness~\cite{xu2025rollouts,zhou2026geoalign,li2026staleflow,hu2025snapshot,li2026pace}. Xu et al.~\cite{xu2025rollouts} preserve reward diversity while reducing the number of rollouts used for policy updates. PACE~\cite{li2026pace} instead couples pool occupancy with effective staleness: excess backlog determines how many trajectories to reject, while policy lag accumulated during waiting and generation, adjusted using prefix-probability changes, informs which to discard. This avoids rejecting a long trajectory solely because generation spans several policy versions, while controlling accumulation in the ready pool.

\paragraph{Experience Replay and Retirement.}

Replay reuses trajectories while controlling their declining relevance as the policy changes~\cite{arnal2026replay,yoo2026replay,zhang2025toolr1,park2026headroom,sun2025improving}. Sun et al.~\cite{sun2025improving} combine difficulty-targeted online selection with replay from a bounded recent buffer, using stored behavior probabilities for importance correction. Selection reduces the steps needed for improvement, while replay reduces fresh generation per step. Headroom-Drift Replay~\cite{park2026headroom} separates learning-value ranking from policy-compatibility checks. These approaches connect reuse decisions to both the remaining value of experience and its compatibility with the current learner.

\section{Reward Computation}
\label{sec:reward-computation}

Reward computation evaluates generated responses or interaction trajectories and produces the rewards used for policy updates. Feedback may come from reward models, LLM judges, executable tests, or rule-based checks. These sources introduce different operational concerns: learned evaluators can produce biased or uncertain judgments, verification procedures can fail or accept outputs that do not satisfy the intended requirements, and combining or assigning feedback can distort the training objective. LLM RLOps therefore manages both the reliability of evaluation and the transformation of evaluation results into training rewards.

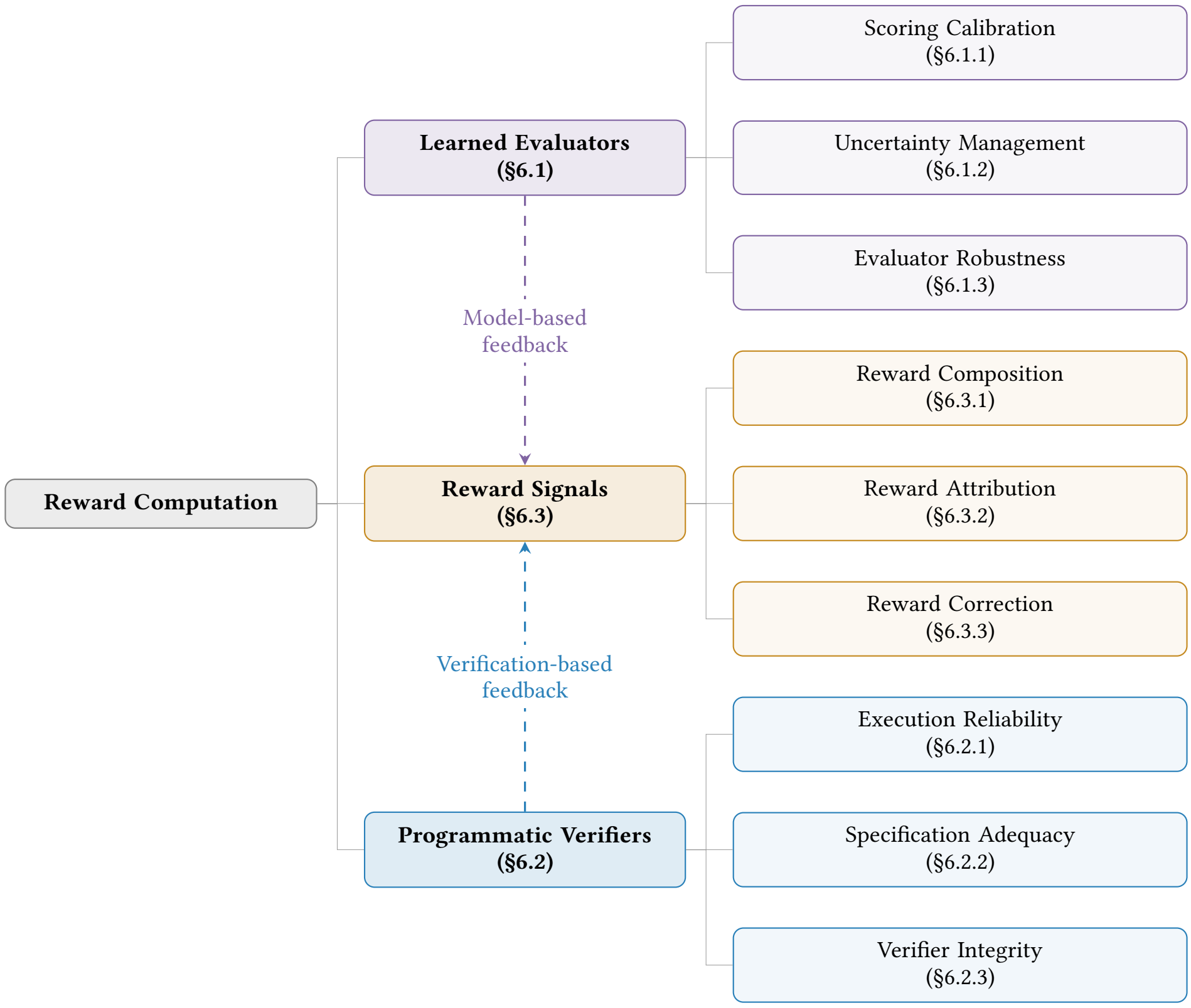
\begin{figure}[htbp]
	\centering
	\begingroup
	\definecolor{rewardModel}{HTML}{8064A2}
	\definecolor{rewardSignal}{HTML}{C58A20}
	\definecolor{rewardVerifier}{HTML}{2980B9}
	
	\tikzset{
		reward node/.style={
			draw,
			thin,
			align=center,
			rounded corners=3pt,
			inner sep=4pt,
			text=black
		},
		reward root/.style={
			reward node,
			draw=gray,
			fill=gray!15,
			text width=3.0cm,
			font=\bfseries\scriptsize
		},
		reward category/.style={
			reward node,
			text width=3.1cm,
			font=\bfseries\scriptsize
		},
		reward leaf/.style={
			reward node,
			text width=4.5cm,
			font=\scriptsize
		},
		feedback relation/.style={
			->,
			>=stealth,
			dashed,
			semithick
		},
		relation label/.style={
			fill=white,
			inner sep=3pt,
			align=center,
			font=\scriptsize,
			text width=2.6cm
		}
	}
	
	\begin{forest}
		for tree={
			anchor=west,
			parent anchor=east,
			child anchor=west,
			grow'=east,
			l sep=14pt,
			s sep=12pt,
			fit=tight,
			calign=midpoint,
			edge={ultra thin, draw=gray!80},
			edge path={
				\noexpand\path [draw,\forestoption{edge}]
				(!u.parent anchor) -- +(6pt,0) |- (.child anchor)
				\forestoption{edge label};
			}
		}
		[Reward Computation, reward root
		[Learned Evaluators\\(\cref{sec:learned-evaluators}),
		reward category,
		name=evaluators,
		draw=rewardModel,
		fill=rewardModel!15
		[Scoring Calibration\\(\cref{sec:scoring-calibration}),
		reward leaf,
		draw=rewardModel,
		fill=rewardModel!6]
		[Uncertainty Management\\(\cref{sec:uncertainty-management}),
		reward leaf,
		draw=rewardModel,
		fill=rewardModel!6]
		[Evaluator Robustness\\(\cref{sec:evaluator-robustness}),
		reward leaf,
		draw=rewardModel,
		fill=rewardModel!6]
		]
		[Reward Signals\\(\cref{sec:reward-signals}),
		reward category,
		name=signals,
		draw=rewardSignal,
		fill=rewardSignal!15
		[Reward Composition\\(\cref{sec:reward-composition}),
		reward leaf,
		draw=rewardSignal,
		fill=rewardSignal!6]
		[Reward Attribution\\(\cref{sec:reward-attribution}),
		reward leaf,
		draw=rewardSignal,
		fill=rewardSignal!6]
		[Reward Correction\\(\cref{sec:reward-correction}),
		reward leaf,
		draw=rewardSignal,
		fill=rewardSignal!6]
		]
		[Programmatic Verifiers\\(\cref{sec:programmatic-verifiers}),
		reward category,
		name=verifiers,
		draw=rewardVerifier,
		fill=rewardVerifier!15
		[Execution Reliability\\(\cref{sec:verification-execution-reliability}),
		reward leaf,
		draw=rewardVerifier,
		fill=rewardVerifier!6]
		[Specification Adequacy\\(\cref{sec:specification-adequacy}),
		reward leaf,
		draw=rewardVerifier,
		fill=rewardVerifier!6]
		[Verifier Integrity\\(\cref{sec:verifier-integrity}),
		reward leaf,
		draw=rewardVerifier,
		fill=rewardVerifier!6]
		]
		]
		\draw[feedback relation, draw=rewardModel]
		(evaluators.south)
		-- node[midway, relation label, text=rewardModel]
		{Model-based\\feedback}
		(signals.north);
		\draw[feedback relation, draw=rewardVerifier]
		(verifiers.north)
		-- node[midway, relation label, text=rewardVerifier]
		{Verification-based\\feedback}
		(signals.south);
	\end{forest}
	\endgroup
	
	\caption{Taxonomy of Reward Computation in LLM RLOps. Solid edges indicate classification relationships. Dashed arrows show feedback contributions from learned evaluators and programmatic verifiers to training reward signals. The two feedback sources may be used independently or combined.}
	\label{fig:reward-computation-taxonomy}
	\Description{A taxonomy tree with three colored branches: Learned Evaluators, Reward Signals, and Programmatic Verifiers. Each branch contains three operational topics with section references. Dashed arrows connect Learned Evaluators and Programmatic Verifiers to Reward Signals.}
\end{figure}

As illustrated in Figure~\ref{fig:reward-computation-taxonomy}, we organize this stage into three categories. \textbf{Learned Evaluators} cover the calibration, uncertainty, and robustness of model-based scoring. \textbf{Programmatic Verifiers} cover the execution reliability, specification adequacy, and integrity of executable checks. These two categories represent complementary feedback sources that may be used independently or together. \textbf{Reward Signals} cover how their outputs are combined across objectives, assigned to responses or individual steps, and corrected when feedback is unreliable. Together, the three categories connect the management of feedback sources with the construction of rewards supplied to policy optimization.

\subsection{Learned Evaluators}
\label{sec:learned-evaluators}

Learned evaluators include preference reward models, process reward models, and LLM judges. They assess properties that are difficult to express through executable checks, such as helpfulness or reasoning quality. Their reliability requires managing three distinct concerns: systematic biases in scoring, uncertainty in individual judgments, and vulnerability to distribution shifts or adversarial outputs. We organize the corresponding techniques into scoring calibration, uncertainty management, and evaluator robustness.

\subsubsection{Scoring Calibration}
\label{sec:scoring-calibration}

Scoring calibration identifies and corrects systematic discrepancies between predicted rewards and the qualities they should measure. Length-bias controls separate content from length-related predictions or remove estimated bias after scoring~\cite{chen2024odin,huang2024posthoc}. Calibration assessment must also reflect how scores are used. AgentOmnia~\cite{jiang2026agentomnia} evaluates candidate reward models through within-group ranking consistency, advantage-sign errors, and advantage-magnitude consistency against reference judgments. These checks expose evaluators that appear adequate under aggregate scoring metrics but distort the direction or relative strength of group-based updates.

\subsubsection{Uncertainty Management}
\label{sec:uncertainty-management}
Uncertainty management controls the influence of poorly supported evaluator judgments. Ensemble approaches estimate uncertainty from disagreement and moderate rewards accordingly~\cite{coste2023reward,zhang2024efficientensemble}. Coste et al.~\cite{coste2023reward} show that conservative objectives based on unfavorable ensemble predictions or uncertainty-weighted rewards reduce overoptimization. Since full-model ensembles increase memory and inference costs, Zhang et al.~\cite{zhang2024efficientensemble} use linear-layer and LoRA-based ensembles. These approaches balance the reliability of reward estimates against the cost of producing them at training scale.

\subsubsection{Evaluator Robustness}
\label{sec:evaluator-robustness}

Evaluator robustness concerns the continued reliability of scoring as the policy generates new outputs or exploits weaknesses in the evaluator. Unlike calibration, which corrects identifiable scoring biases, robustness addresses changes in the conditions under which the evaluator operates. Existing techniques include weight averaging, iterative evaluator updates, and adversarial repair~\cite{bai2022training,rame2024warm,wolf2025overoptimisation,liang2026ares}.

WARM~\cite{rame2024warm} averages independently fine-tuned reward models in parameter space to improve robustness to distribution shifts and inconsistent preferences, while retaining a single model for scoring. Across successive RLHF iterations, Wolf et al.~\cite{wolf2025overoptimisation} examine how feedback retention, reward-model updates, and policy initialization affect overoptimization. Their findings indicate that refreshing the evaluator can help, but recovery also depends on how the policy and its training history are managed.

Adversarial approaches explicitly search for outputs that reveal evaluator weaknesses. ARES~\cite{liang2026ares} identifies cases in which both the policy and reward model fail, repairs the reward model using the discovered examples, and then uses the improved evaluator to optimize the policy. These approaches extend evaluator management beyond initial validation: scoring reliability must be reassessed as the policy and its output distribution evolve.

\subsection{Programmatic Verifiers}
\label{sec:programmatic-verifiers}

Programmatic verifiers evaluate outputs through executable tests, parsers, rule checks, or formal verification. Their reliability depends on three conditions: the checking procedure must execute correctly, its acceptance conditions must adequately represent the task, and the policy must not be able to manipulate the verification mechanism. These conditions motivate the distinction between execution reliability, specification adequacy, and verifier integrity.

\subsubsection{Execution Reliability}
\label{sec:verification-execution-reliability}

Execution reliability concerns whether a verification attempt produces a valid result under the intended execution conditions. In coding and agent training, verification depends on software dependencies, process execution, resource limits, and external services. A failed verification attempt can therefore reflect an infrastructure problem rather than an incorrect policy output. Relevant systems work supports controlled execution and preserves the information needed to interpret these outcomes~\cite{huang2026dsec,agentcpm2026explore}.

DSec~\cite{huang2026dsec} provides isolated execution environments and exposes command outputs and return status used in reward computation. AgentCPM-Explore~\cite{agentcpm2026explore} addresses the downstream consequences of unreliable execution by filtering trajectories whose outcomes cannot be meaningfully attributed to the agent, including those affected by environmental disruptions. These studies motivate distinguishing a valid negative verdict from an unsuccessful verification attempt. The focus here is the validity of the evidence used for scoring; the availability and recovery of interaction services are discussed under Environment Reliability.

\subsubsection{Specification Adequacy}
\label{sec:specification-adequacy}

Specification adequacy examines whether the verifier's acceptance conditions capture the intended task. Even a correctly executed check can accept an undesirable solution when tests omit relevant cases or formal specifications leave important requirements unconstrained. Studies in reasoning and program verification demonstrate that such gaps can become exploitable training incentives~\cite{helff2026gaming,tan2026formal}.

Helff et al.~\cite{helff2026gaming} show that verification based only on observed instances can reward enumeration strategies instead of generalizable rule induction. They introduce isomorphic verification to check whether outputs remain valid under logically equivalent task transformations, exposing shortcuts admitted by the original checks. In formal verification, Tan~\cite{tan2026formal} reports that weak specifications allow models to increase verified success without implementing the intended solutions, motivating the removal of underspecified and vulnerable tasks. These findings establish a shared management concern: acceptance criteria must be examined for what they fail to enforce, rather than treated as complete definitions of correctness.

\subsubsection{Verifier Integrity}
\label{sec:verifier-integrity}

Verifier integrity protects the checking mechanism and its evidence from policy-controlled interference. Agents that can edit files, execute programs, or access internal services may alter tests, replace verification logic, or obtain information through unintended channels. Existing work investigates both the emergence of these behaviors and mechanisms that restrict the agent's access to the scoring infrastructure~\cite{khalifa2026countdown,huang2026dsec,wang2026catch}.

Countdown-Code~\cite{khalifa2026countdown} provides a controlled setting for studying reward hacking through writable tests and verification logic. It supports examining how these behaviors emerge and how tampering detection or penalties affect training. DSec~\cite{huang2026dsec} implements file, socket, and network access controls that limit access to residual answers and internal communication channels. Whereas specification adequacy concerns weaknesses in the original acceptance conditions, verifier integrity concerns whether the policy can interfere with the mechanism that applies those conditions.

\subsection{Reward Signals}
\label{sec:reward-signals}

Reward signals are the values supplied to policy optimization after evaluation. Their construction determines how different objectives influence learning, which parts of a trajectory receive credit, and how unreliable feedback is handled. We distinguish reward composition, reward attribution, and reward correction according to these three decisions. This perspective also captures failures that arise even when individual feedback sources operate as designed.

\subsubsection{Reward Composition}
\label{sec:reward-composition}

Reward composition determines how objectives with different scales and reliability influence training~\cite{dai2023saferlhf,han2025user}. Safe RLHF~\cite{dai2023saferlhf} separates helpfulness rewards from safety costs and balances them through constrained optimization. Such separation makes trade-offs explicit, but adding a component can also create an unintended incentive.

Sun and Kazakov~\cite{sun2026peak} report that rewarding the appearance of retrieved entities in answers admits a shortcut in knowledge-graph tool use: issuing one call, copying an entity, and stopping. Yao et al.~\cite{yao2026multimodal} likewise find that an added visual-evidence reward worsens hacking under keyword-based verification, while the corresponding intervention helps in a setting with a VLM judge. These cases motivate validating the behavior induced by a reward combination, including the reliability of each component's verifier.

\subsubsection{Reward Attribution}
\label{sec:reward-attribution}
Reward attribution determines which tokens, steps, or trajectory segments receive credit or penalties. Its operational concern is whether the assignment and accumulation of feedback create incentives that depart from the intended task. PURE~\cite{cheng2025pure} demonstrates that summing process rewards can encourage exploitation of highly rewarded intermediate steps. Its minimum-based formulation changes how process scores determine learning credit to reduce this failure. This evidence motivates checking the behavior induced by reward aggregation, rather than assuming that finer-grained feedback necessarily provides more reliable supervision.

\subsubsection{Reward Correction}
\label{sec:reward-correction}

Reward correction addresses misleading feedback from verifier errors, flawed reasoning, or environmental failures~\cite{cai2025noisy,ding2025fapo,agentcpm2026explore}. Cai et al.~\cite{cai2025noisy} model verifier feedback as an asymmetric noisy channel and correct its policy-gradient bias. For research-agent training with simulated feedback, WMRL~\cite{yang2026worldmodels} retains a small stream of real executions to debias world-model rewards and weights gradient estimates according to their variance. These corrections support cheaper feedback without treating it as equivalent to direct execution. When an outcome cannot be reliably attributed to the agent, filtering it from learning remains an alternative~\cite{agentcpm2026explore}.

\section{Policy Optimization}
\label{sec:policy-optimization}

Policy optimization uses collected trajectories and their rewards to update the model parameters, producing new policy versions for subsequent experience collection. Operating this process requires coordinating gradient computation, memory usage, and communication across training devices. It also requires accounting for differences between the policy that generated the trajectories and the policy being updated, monitoring whether updates sustain effective learning, and preserving training progress against execution failures.

\begin{figure}[htbp]
	\centering
	\includegraphics[width=\textwidth]{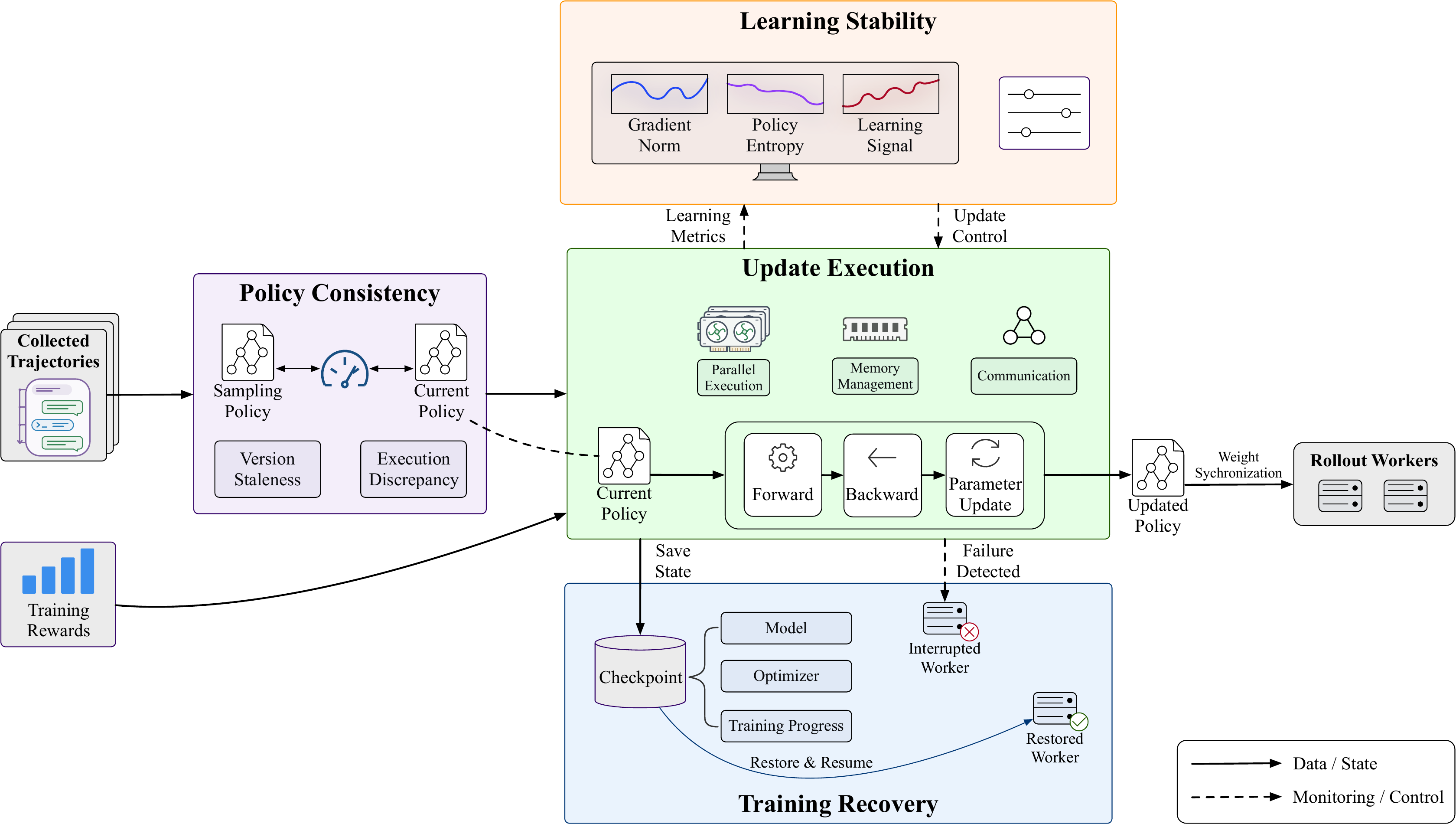}
	\caption{Overview of Policy Optimization in LLM RLOps. Update Execution transforms trajectories and rewards into updated policy parameters. Policy Consistency manages differences between sampling and training policies, Learning Stability monitors and controls learning dynamics, and Training Recovery restores execution from saved training state after failures.}
	\label{fig:policy-optimization}
\end{figure}

As illustrated in Figure~\ref{fig:policy-optimization}, we organize this stage into four complementary responsibilities. \textbf{Update Execution} manages the computation and resources needed to perform parameter updates. \textbf{Policy Consistency} addresses policy staleness and discrepancies between rollout and training execution, constraining or correcting their effects on learning. \textbf{Learning Stability} monitors gradients, policy entropy, and effective learning signals to detect and mitigate unstable or degenerate updates. \textbf{Training Recovery} manages recoverable training state and restores execution after runtime failures. These responsibilities operate around the same learner: consistency management connects collected experience to the current policy, stability management controls learning behavior during updates, and recovery management preserves continuity when execution is interrupted.

\subsection{Update Execution}
\label{sec:update-execution}

Update execution organizes the computation and resources required to perform policy updates. LLM reinforcement learning combines multiple model roles, changing sequence lengths, and repeated exchanges between training and rollout workers. Existing work addresses these demands through execution orchestration, memory management, computation optimization, and parameter transfer.

\paragraph{Execution Orchestration.}

Execution orchestration maps RL workflows to model operations and device assignments~\cite{yao2023deepspeed,havrilla2023trlx,xiao2023adaptive}. Systems separate workflow control from distributed execution and coordinate model roles across shared devices~\cite{sheng2025hybridflow,mei2024real,shen2024nemo,wu2025llamarl,wang2025distflow,yu2025rlinf,zhong2024stagefusion}. HybridFlow~\cite{sheng2025hybridflow}, for example, combines workflow-level control with distributed model execution. Further work adapts execution to heterogeneous accelerators, distributed dataflow, and changing runtime configurations~\cite{yan2025arealhex,he2025hetrl,feng2025mindspeed,hu2026jigsawrl,jiang2026nereus}.

Orchestration also extends across concurrent runs. MARLaaS~\cite{yu2026marlaas} shares a base model through task-specific LoRA adapters and overlaps different tasks' training, rollout, and interaction phases while preserving policy-version consistency within each task. PlexRL~\cite{zhang2026plexrl} schedules model operations as services and packs work from multiple jobs into available cluster capacity. These designs broaden resource coordination beyond optimizing the phases of a single run.

\paragraph{Memory Management.}
Memory management reduces the simultaneous residency of model parameters, optimizer states, and activations. Existing techniques include sharing parameters across model roles, updating lightweight adapters, and moving inactive state out of GPU memory~\cite{santacroce2023efficient,sidahmed2024parameter,shen2024nemo,wu2025llamarl}. Santacroce et al.~\cite{santacroce2023efficient} combine model sharing with dynamically enabled LoRA modules to reduce the memory requirements of PPO. Sidahmed et al.~\cite{sidahmed2024parameter} evaluate parameter-efficient reward modeling and policy training across several tasks, showing that reduced trainable state can lower memory use and training time while retaining competitive performance. These methods address persistent model-state costs, complementing techniques that reduce transient activation costs during execution.

\paragraph{Computation Optimization.}
Computation optimization exploits structure in the update workload to reduce repeated work and device imbalance. One line of research reuses shared trajectory prefixes during forward and backward computation~\cite{wang2025tree,zhang2026arealdta,li2026prefixflow}. Tree Training~\cite{wang2025tree} expresses branch-wise training as a weighted computation over shared tokens, allowing common prefixes to be processed once while preserving the intended calculation. A second line uses information already available from rollouts to anticipate expert workloads in MoE training~\cite{jin2026relibra,zhou2026foremoe}. ReLibra~\cite{jin2026relibra}, for example, uses recorded routing decisions to guide expert reordering and replication at different execution timescales. The two lines exploit complementary properties of RL workloads: repeated input structure and advance knowledge of the computation that training will perform.

\paragraph{Parameter Transfer.}
Parameter transfer distributes updated weights to rollout workers across potentially different layouts and network conditions. Existing approaches optimize layout conversion and transfer topology, reuse available weight replicas, or transmit only relevant parameter changes~\cite{jiang2026weightbridge,ye2026tensorhub,miahi2026sparsity,ruan2026aurorarl}. WeightBridge~\cite{jiang2026weightbridge} derives the correspondence between training and rollout layouts and plans load-balanced transfers without redundant movement. TensorHub~\cite{ye2026tensorhub} instead exposes versioned weights through a reference-oriented storage abstraction that uses existing GPU-resident replicas to serve requests. These approaches reduce the cost of making a new policy available; the learning consequences of delayed or numerically different policies are addressed separately below.

\subsection{Policy Consistency}
\label{sec:policy-consistency}

Policy consistency manages differences between the policy that generated experience and the policy used to compute updates. Such differences arise from intervening parameter updates, numerical implementations, and changes in the model's execution path. The objective is to constrain or compensate for consequential discrepancies while retaining the efficiency benefits of asynchronous and specialized execution.

\paragraph{Policy Staleness.}
Policy staleness accumulates while trajectories are generated, transferred, or stored. Systems combine synchronization, bounded delay, and off-policy correction to manage its effects~\cite{noukhovitch2024asynchronous,fu2026areal,xiao2025echo,tian2026mugrpo,lu2025periodic}. AReaL~\cite{fu2026areal} uses staleness-aware PPO, whereas $\mu$-GRPO~\cite{tian2026mugrpo} permits more reuse through relaxed clipping and selective suppression of destabilizing suffix updates. Periodic Asynchrony~\cite{lu2025periodic} instead coordinates generation and weight synchronization to preserve on-policy updates. Other studies examine synchronization intervals, learning rates, temperature, and historical probabilities~\cite{song2026collapse,pan2026cool,guan2026missing}. Acceptable delay therefore depends on both the update rule and the information retained with experience.

\paragraph{Numerical Consistency.}

Identical parameter versions can produce different probabilities under different precisions, kernels, or reduction orders~\cite{qi2025fp16,zhang2025deterministic,zhong2026mismatch}. MiniMax-M1~\cite{minimax2025m1} reports a training--inference probability mismatch that prevented reward growth. Layer-wise diagnosis localized the main error to high-magnitude activations in the output head; computing that head in FP32 reduced the discrepancy and restored learning progress.

Quantized rollout systems address related discrepancies by aligning precision flows or correcting their effects during updates~\cite{xi2026jetrl,qiu2026fp8rl,gu2026qarl,li2026qurl,zhuge2026quads}. Jet-RL~\cite{xi2026jetrl} aligns precision across training and generation, whereas QuRL~\cite{li2026qurl} adapts clipping to quantization-induced mismatch. The appropriate intervention therefore depends on where the discrepancy originates and how it enters the update calculation.

\paragraph{Execution-Path Consistency.}
Execution-path consistency concerns differences in the experts, context, or generated token stream used during rollout and training. Relevant work preserves MoE routing decisions, reconciles agent-harness outputs with training inputs, and corrects differences introduced by sparse rollout execution~\cite{ma2025routers,du2026legorl,luo2026sparserl,zhou2026sparrow}. Rollout Routing Replay~\cite{ma2025routers} records expert selections during generation and reuses them during training, reducing divergence caused by inconsistent routing. Sparse-RL~\cite{luo2026sparserl} addresses a different discrepancy: a sampler using compressed context can differ from both the previous dense policy and the current learner. It combines rejection sampling with importance-based reweighting to manage this mismatch. These studies show why matching model versions and numerical precision is insufficient when the computation producing the sampled behavior also changes.

\subsection{Learning Stability}
\label{sec:learning-stability}

Learning stability concerns whether repeated updates sustain useful learning without excessive policy drift, exploration collapse, or ineffective gradients. Existing studies connect observations of training dynamics with interventions on update magnitude, behavioral diversity, and the construction of optimization signals. The emphasis is on managing the running learning process, including failures that occur despite valid trajectories and rewards.

\paragraph{Update Regulation.}

Update regulation controls parameter changes through reward baselines, regularization, reference-policy management, and regression-aware intervention~\cite{hou2024chatglm,liu2025prolonged,cagatan2025failure,lin2026selfregress}. These controls must account for execution errors: Dong et al.~\cite{dong2026probing} link training--inference discrepancies to biased optimization and find that stronger token-level clipping can accelerate the discrepancy's growth in their experiments. A nominally stabilizing setting can therefore interact adversely with system-induced bias.

Diagnosis complements these controls through failure reproduction, runtime monitoring, and automated remediation~\cite{wang2026moeproxy,zhang2026failure}. In multi-agent workflows, Zeng et al.~\cite{zeng2026workflows} identify role-specific drift and dominance of one role within a shared policy that aggregate metrics can conceal. They motivate tracking per-role gradients, perplexity, and divergence alongside overall accuracy. These observations help determine whether intervention should target a global update setting or a particular role.

\paragraph{Exploration Preservation.}
Exploration preservation addresses the concentration of policy behavior before useful learning is complete. Studies examine entropy collapse, repetitive agent behavior, and the loss of solution diversity under continued optimization~\cite{yu2025dapo,wang2025ragen,yuan2026diversity}. DAPO~\cite{yu2025dapo} uses asymmetric clipping to give low-probability tokens more room to increase their probability, mitigating premature entropy collapse. Yuan et al.~\cite{yuan2026diversity} identify a complementary failure mechanism: repeated updates on already-saturated problems can concentrate probability on familiar solutions while reducing broader solution coverage. Their analysis motivates restricting updates according to remaining learning value. These findings distinguish preserving local response diversity from preventing overtraining that narrows the range of problems or solutions a policy can cover.

\paragraph{Learning Signal Preservation.}
Learning signal preservation ensures that available rewards produce informative and appropriately scaled gradients. Group-based updates can lose their signal when rewards are homogeneous, while normalization can distort signals across heterogeneous agents or reward channels~\cite{he2026advantage,feng2026drmas,wang2026darkroom,wang2026ragen2,pu2026explore}. He et al.~\cite{he2026advantage} monitor the prevalence of ineffective gradients through an advantage-collapse metric and introduce virtual reward samples to restore learning from homogeneous groups. Dr. MAS~\cite{feng2026drmas} addresses scale distortion by normalizing advantages using each agent's own reward statistics, reducing gradient instability caused by a shared baseline. These methods manage how feedback becomes an optimization signal, complementing the preceding chapter's treatment of whether the feedback itself is valid.

\subsection{Training Recovery}
\label{sec:training-recovery}

Training recovery preserves valid progress and restores execution after process failures, device loss, or resource preemption. Because learners and rollout workers maintain different states and may advance asynchronously, recovery must coordinate saved state with surviving components. Existing systems address these requirements through checkpoint management and fault isolation.

\paragraph{Checkpoint Management.}
Checkpoint management trades saving overhead against progress lost after interruption. WeChat-YATT~\cite{wu2025wechat} uses asynchronous checkpointing to increase saving frequency and attempts an additional checkpoint when training resources must be reclaimed for online services. Its distributed checkpoints and data-consumption state support resumption on clusters with different GPU counts. This design connects recovery to resource availability: preserving progress is useful only if the saved state can be restored under the allocation available after interruption.

\paragraph{Fault Isolation.}

Fault isolation limits disruption to failed roles or workers and coordinates their return~\cite{chen2025role,ruan2026aurorarl,sheng2026laminar}. RobustRL~\cite{chen2025role} uses role-aware monitoring and a detect--restart--reconnect procedure so unaffected rollout workers can continue generation during trainer recovery. Laminar~\cite{sheng2026laminar} separates rollout execution from the learner and parameter-relay service. In an injected rollout-machine failure, it restores generation capacity by provisioning a replacement while the learner continues using surviving workers. This evidence concerns worker recovery; restoring a failed learner additionally requires its training state and data-consumption position.

\section{Model Evaluation}
\label{sec:model-evaluation}

Model evaluation assesses whether LLM reinforcement learning produces the intended capability gains and identifies regressions or risks introduced by training. During an RL run, successive checkpoints must be evaluated through procedures that produce comparable measurements and reveal changes in model behavior. These procedures can themselves yield misleading evidence. For example, Hochlehnert et al.~\cite{hochlehnert2025sober} show that reasoning scores are sensitive to decoding parameters, prompt formatting, and software configurations, while Harmon et al.~\cite{harmon2025forgetting} demonstrate that aggregate accuracy can conceal the loss of previously correct responses. Such problems can lead to retaining checkpoints with overstated gains or undetected regressions. Managing this stage therefore involves controlling evaluation configurations, checking measurement reproducibility, and maintaining regression audits across model versions. These activities make evaluation evidence dependable enough to support checkpoint selection and change acceptance.

\begin{figure}[htbp]
	\centering
	\includegraphics[width=\textwidth]{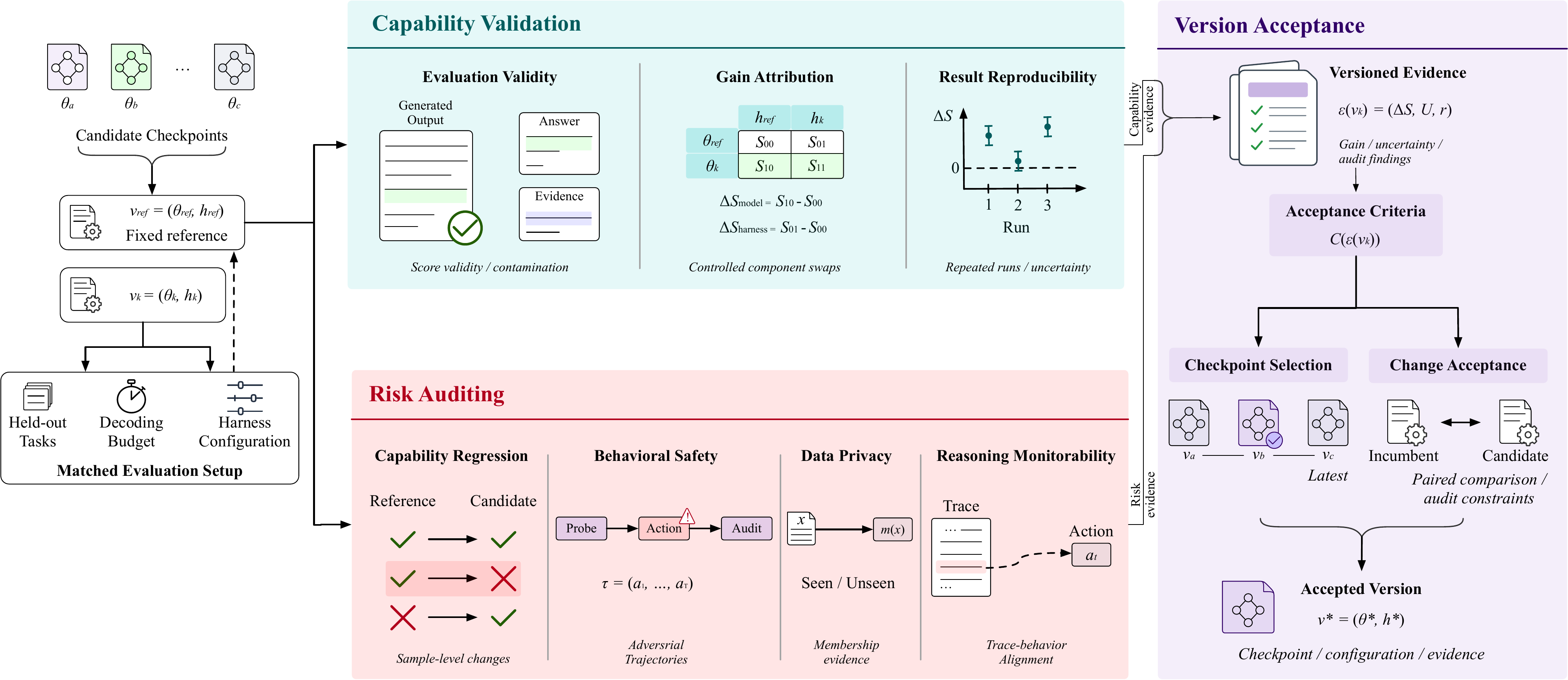}
	\caption{Overview of Model Evaluation in LLM RLOps. Capability Validation and Risk Auditing provide complementary evidence about candidate versions. Version Acceptance uses this evidence for checkpoint selection and change acceptance, with accepted versions providing references for subsequent evaluation.}
	\label{fig:model-evaluation}
\end{figure}

As illustrated in Figure~\ref{fig:model-evaluation}, we organize this stage into three complementary categories. \textbf{Capability Validation} examines whether measured improvements are valid, attributable to the changes being evaluated, and reproducible. \textbf{Risk Auditing} examines capability regressions, unsafe behavior, data exposure, and limitations in reasoning monitorability. These two branches provide complementary evidence about gains and risks, using matched evaluation conditions when comparing candidate versions against a reference. \textbf{Version Acceptance} uses this evidence to select checkpoints and determine whether proposed model or configuration changes should be retained. An accepted version can then serve as the reference for subsequent evaluation, connecting individual assessments across successive training iterations.

\subsection{Capability Validation}
\label{sec:capability-validation}

Capability validation establishes whether evaluation results provide credible evidence of improvement after RL. A reported gain must measure the intended capability, be attributable to the intervention being assessed, and remain reliable across repeated comparisons. Existing work addresses these requirements through evaluation validity, gain attribution, and result reproducibility.

\paragraph{Evaluation Validity.}
Evaluation validity checks whether scores support the claimed capability and whether evaluation data remain independent of training~\cite{wang2026visualutility,tao2025contamination,zhang2026structural,maniyar2026measuring}. Wang~\cite{wang2026visualutility} uses image interventions to separate answer accuracy from reliance on visual evidence. Trace analysis additionally distinguishes locally coherent steps from globally valid reasoning~\cite{samineni2025coherence}. Position-confounded RL experiments show how high rewards can reflect a learned answer-position shortcut~\cite{maniyar2026measuring}. For contamination, Tao et al.~\cite{tao2025contamination} probe entropy changes under self-critique to identify RL training examples. These checks test what a score measures before interpreting it as generalization.

\paragraph{Gain Attribution.}
Gain attribution separates policy learning from changes in encountered states, supporting components, or the training pipeline~\cite{liu2026reach,chen2026agent2,wang2026visualutility}. Liu and Qian~\cite{liu2026reach} clone states reached by one policy and let another continue, separating the contributions of reaching and solving. Chen et al.~\cite{chen2026agent2} evaluate post-training engineering agents under fixed budgets and inspect their execution artifacts, revealing whether successful workflows actually use RL or primarily supervised adaptation. Such controls identify which intervention accounts for a measured gain.

\paragraph{Result Reproducibility.}
Reproducibility checks whether gains persist across repeated runs and controlled evaluation settings~\cite{hochlehnert2025sober,ma2025rethinking,liang2026dataflex}. Liang et al.~\cite{liang2026dataflex} compare data-selection policies using a common GRPO recipe, matched seeds, and multiple benchmarks. The tested policies do not establish reproducible gains over uniform training in their settings, and benchmark composition changes method rankings. These findings motivate preserving configurations and per-run results, reporting uncertainty, and checking sensitivity to the evaluation mixture.

\subsection{Risk Auditing}
\label{sec:risk-auditing}

Risk auditing examines adverse changes that improvements on the target task may conceal. Its scope includes losses in existing capabilities, undesirable behavior, exposure of training-data membership, and reduced visibility into model behavior. These concerns require distinct audit procedures because a model can pass one type of assessment while failing another.

\paragraph{Capability Regression.}
Regression audits track capabilities that adaptation should preserve, including knowledge, language fidelity, and useful output diversity~\cite{harmon2025forgetting,dobler2026grpo,karouzos2026diversity,liu2026efficiency,park2025crosslingual}. Harmon et al.~\cite{harmon2025forgetting} separate correct-to-incorrect transitions from newly acquired successes, exposing forgetting hidden by aggregate accuracy. Karouzos et al.~\cite{karouzos2026diversity} distinguish removal of incorrect outputs from narrowing among correct ones. Cross-lingual collapse further shows why rising reasoning scores should be assessed alongside language consistency~\cite{park2025crosslingual}. These audits preserve evidence of trade-offs that a single score can conceal.

\paragraph{Behavioral Safety.}
Behavioral safety audits investigate whether post-training introduces undesirable objectives or behaviors across interaction settings~\cite{marks2025auditing,macdiarmid2025natural,elcock2026alignment}. Marks et al.~\cite{marks2025auditing} construct a model with a hidden objective and evaluate whether auditors can uncover it through behavioral probing, interpretability, and training-data analysis. Their approach tests the effectiveness of the auditing procedure against a known failure. MacDiarmid et al.~\cite{macdiarmid2025natural} show that reward hacking learned in coding environments can generalize to broader misaligned behavior in their experimental setting, motivating audits beyond the tasks used for training. The resulting evidence helps determine whether a candidate version preserves acceptable behavior across contexts. Here, the concern is the behavior acquired by the model; the reliability of the reward-producing components is examined under Reward Computation.

\paragraph{Data Privacy.}
Privacy audits test whether RL reveals training-set membership~\cite{liu2025membership,zhang2026structural,tao2025contamination}. Liu et al.~\cite{liu2025membership} compare a model with its initial checkpoint using changes in verifiable success and response distributions across multiple rollouts. Other methods detect structural convergence in reasoning traces without the initial checkpoint~\cite{zhang2026structural}. These signals support privacy and contamination audits, although evidence of membership does not itself establish extraction of the underlying data.

\paragraph{Reasoning Monitorability.}
Reasoning monitorability audits assess whether observable reasoning traces retain information useful for detecting influences on model behavior~\cite{xiong2026monitorability,little2026length}. Xiong et al.~\cite{xiong2026monitorability} find that monitorability gains during RLVR depend on the training data and do not follow automatically from improved task performance. Little~\cite{little2026length} studies a complementary failure mode: length-penalized RL can leave accuracy largely intact while reducing the disclosure of hints that influence the model's answers. A checkpoint can therefore become less observable even when conventional performance metrics remain stable. These findings motivate evaluating monitorability separately when changing training objectives or compression settings, using interventions that test the relationship between visible traces and behavior rather than relying on trace length or fluency.

\subsection{Version Acceptance}
\label{sec:version-acceptance}

Version acceptance uses evaluation evidence to decide which checkpoints and proposed changes should be retained. The candidate may be a saved model from an ongoing run or a model combined with revised instructions, tools, or execution settings. Existing work distinguishes checkpoint selection within a run from change acceptance across successive improvement cycles.

\paragraph{Checkpoint Selection.}
Checkpoint selection weighs evaluation costs against the risk of retaining a weaker model. Existing approaches derive selection signals from training statistics or use independent downstream evaluations to identify deterioration~\cite{nguyen2025checkpoint,zhang2026evalstop}. Nguyen et al.~\cite{nguyen2025checkpoint} identify difficult examples through per-sample uncertainty and average their rewards over a short training window to rank checkpoints without additional forward passes. This reduces selection overhead, while the relationship between the resulting signal and generalization is evaluated empirically. Zhang et al.~\cite{zhang2026evalstop} instead use consecutive declines in downstream evaluation to terminate a run, preserve its best checkpoint, and release its resources. Their scheduling benefits are evaluated in simulation. Together, these approaches connect evaluation with checkpoint retention and training duration, while showing that the most recent checkpoint need not be the preferred output.

\paragraph{Change Acceptance.}
Change acceptance determines whether a proposed model or component revision should replace the current version. Systems that jointly improve policies and their surrounding execution harnesses use controlled comparisons and reversible changes to manage this decision~\cite{qi2026vace,mao2026safeevolve}. Qi et al.~\cite{qi2026vace} compare a proposed harness revision with the incumbent while holding the updated model fixed; only revisions that improve validation performance guide subsequent RL training. Mao et al.~\cite{mao2026safeevolve} apply paired acceptance tests to bounded component changes and retain supporting evidence and rollback metadata. Related systems extend acceptance procedures to autonomous training branches, production model iterations, and joint updates of skills, adapters, and routing configurations~\cite{chen2026evotrainer,nie2026characterflywheel,yao2026mera}. Their mechanisms include backtesting, offline and online comparisons, and replay checks that preserve task quality. Such procedures are especially relevant when models propose their own improvements: a candidate intervention must pass an evaluation step before its effects propagate into subsequent training cycles.

\section{Challenges and Future Directions}
\label{sec:challenges}

\begin{figure}[htbp]
	\centering
	\resizebox{\linewidth}{!}{%
		\begingroup
		\definecolor{rlInk}{HTML}{293744}
		\definecolor{rlGray}{HTML}{687785}
		\definecolor{rlLine}{HTML}{BECAD3}
		\definecolor{rlBlue}{HTML}{526F8A}
		\definecolor{rlPurple}{HTML}{526F8A}
		\definecolor{rlGold}{HTML}{526F8A}
		\definecolor{rlGreen}{HTML}{526F8A}
		\definecolor{rlRed}{HTML}{AE8C62}
		\begin{tikzpicture}[
			font=\fontsize{10}{12}\selectfont,
			text=rlInk,
			line width=0.5pt,
			>={Stealth[length=3.8pt,width=3.2pt]},
			direction/.style={font=\fontsize{10.5}{12.5}\selectfont\bfseries,align=center},
			annotation/.style={font=\fontsize{9}{11}\selectfont,text=rlGray,align=center},
			stage/.style={circle,minimum size=0.40cm,inner sep=0pt,
				font=\fontsize{8}{9}\selectfont,fill=white},
			feedback/.style={line width=0.65pt,dash pattern=on 4pt off 3pt,
				rounded corners=10pt,draw=rlRed!85,->}
			]
			% Equal run cards: one expanded only enough to show stage relationships.
			\fill[fill=rlBlue!5,rounded corners=6pt]
			(0.9,2.15) rectangle (6.05,4.55);
			\draw[draw=rlBlue!65,rounded corners=6pt,line width=0.6pt]
			(0.9,2.15) rectangle (6.05,4.55);
			\fill[fill=rlGray!3,rounded corners=6pt]
			(9.0,2.15) rectangle (14.15,4.55);
			\draw[draw=rlLine,rounded corners=6pt,line width=0.6pt]
			(9.0,2.15) rectangle (14.15,4.55);
			% Observability is a property of the entire run, not an extra stage.
			\node[direction,text=rlInk] at (3.475,4.95)
			{Training Process Observability};
			\draw[draw=rlBlue!65,line width=0.5pt] (3.475,4.74) -- (3.475,4.55);
			\node[font=\fontsize{10.5}{12}\selectfont] at (3.475,4.06)
			{Training run $A$};
			\node[font=\fontsize{10.5}{12}\selectfont] at (11.575,4.06)
			{Training run $B$};
			% The spine denotes stage order; the lower bracket denotes coordination.
			\draw[draw=rlPurple!40,line width=0.55pt] (1.675,3.35) -- (5.275,3.35);
			\draw[draw=rlLine,line width=0.55pt] (9.775,3.35) -- (13.375,3.35);
			\foreach \i/\x in {1/1.675,2/2.575,3/3.475,4/4.375,5/5.275}{
				\node[stage,draw=rlPurple!65,text=rlPurple] (a\i) at (\x,3.35) {\i};
			}
			\foreach \i/\x in {1/9.775,2/10.675,3/11.575,4/12.475,5/13.375}{
				\node[stage,draw=rlGray!55,text=rlGray] at (\x,3.35) {\i};
			}
			\draw[draw=rlPurple!65,rounded corners=3pt,line width=0.5pt]
			(1.675,3.04) -- (1.675,2.88) -- (5.275,2.88) -- (5.275,3.04);
			\draw[draw=rlPurple!65] (3.475,2.88) -- (3.475,2.73);
			\node[direction,text=rlInk] at (3.475,2.49)
			{Cross-Stage Coordination};
			\node[annotation] at (11.575,2.53) {Different training setting};
			% Knowledge transfer links settings rather than adding another stage.
			\node[direction,text=rlInk] at (7.525,4.02) {Generalizable\\RLOps};
			\draw[draw=rlGold!85,<->,line width=0.7pt]
			(6.29,3.35) -- (8.76,3.35);
			\node[annotation,text=rlGray] at (7.525,2.85) {Across settings};
			% A single bracket expresses shared management at platform scope.
			\draw[draw=rlGreen!75,rounded corners=5pt,line width=0.65pt]
			(3.475,2.15) -- (3.475,1.48) -- (11.575,1.48) -- (11.575,2.15);
			\fill[rlGreen] (7.525,1.48) circle (1.4pt);
			\draw[draw=rlGreen!75,line width=0.65pt] (7.525,1.48) -- (7.525,1.10);
			\node[direction,text=rlInk] at (7.525,0.84) {Multi-Run Management};
			\node[annotation,text=rlGray] at (7.525,0.43) {Shared resources};
			% Generational feedback is the outermost relationship.
			\draw[feedback] (14.15,4.02) -- (14.68,4.02) -- (14.68,6.02)
			-- (0.37,6.02) -- (0.37,4.02) -- (0.9,4.02);
			\node[font=\fontsize{12}{14}\selectfont\bfseries,text=rlInk,
			fill=white,inner xsep=10pt,inner ysep=4pt]
			at (7.525,6.02) {RLOps for RSI};
			\node[annotation,text=rlGray] at (7.525,5.57) {Across improvement generations};
		\end{tikzpicture}
		\endgroup
	}
	\caption{Relationships among future directions for LLM RLOps. Within a training run, observability provides evidence for coordinated interventions across stages. Across runs, generalizable RLOps enables knowledge transfer between training settings, while multi-run management coordinates shared resources. Building on these capabilities, RLOps for RSI extends operational management to successive cycles of model-guided improvement. Numbers 1--5 correspond to the five training stages in our taxonomy.}
	\label{fig:rlops-research-directions}
\end{figure}
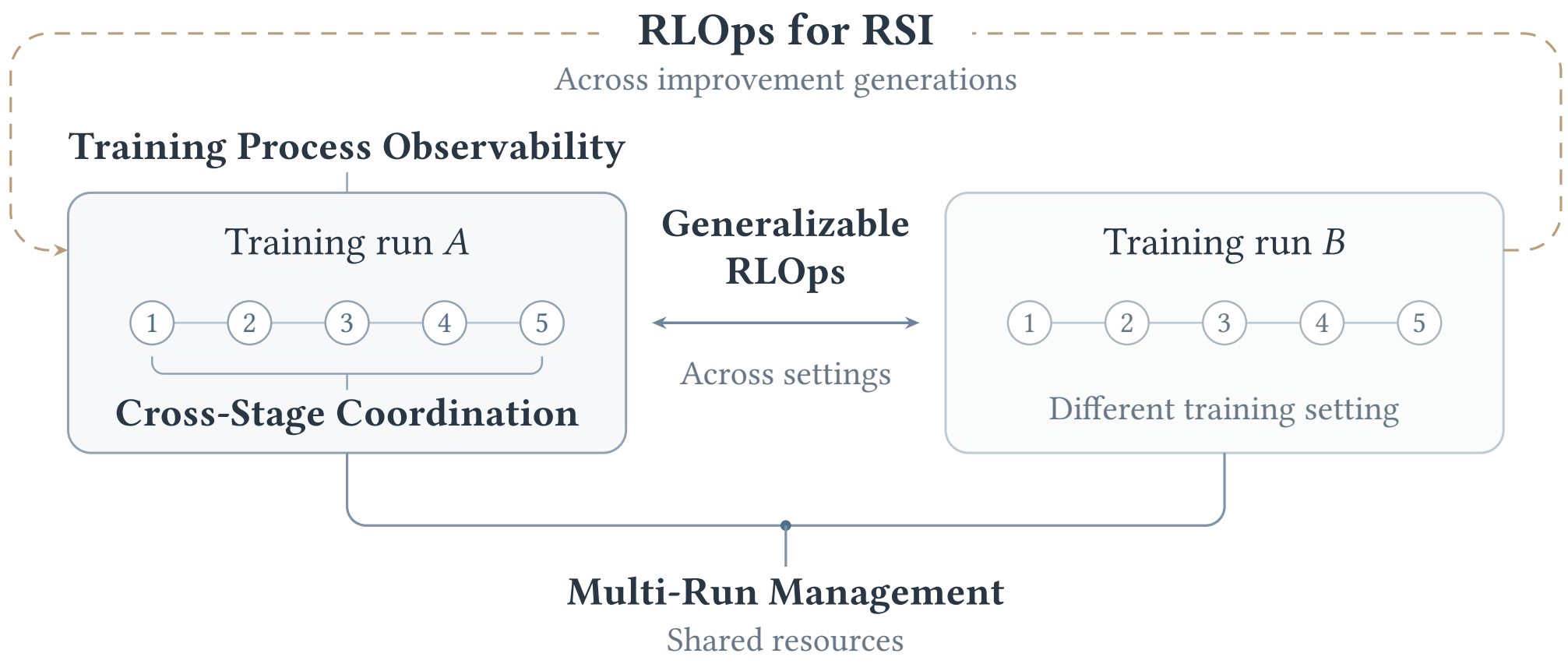

The preceding sections examine operational techniques across five stages of LLM reinforcement learning. Their integration raises broader questions about how to understand a training run, coordinate its management, and extend operational knowledge beyond the setting in which it was developed. As illustrated in Figure~\ref{fig:rlops-research-directions}, we organize these questions into five directions. Training Process Observability and Cross-Stage Coordination concern the management of a complete run. Generalizable RLOps and Multi-Run Management extend this scope to different training settings and shared infrastructure. RLOps for RSI considers how these capabilities can support successive cycles of model-guided improvement.

\subsection{Training Process Observability}
\label{sec:training-process-observability}

Effective management requires understanding whether training is progressing as intended and where intervention is needed. Existing work already goes beyond monitoring a single reward or loss curve. RFT-FM combines multiple training signals to detect failures, diagnose their causes, and guide remediation~\cite{zhang2026failure}. Studies of multi-agent training further show that role-specific gradient dynamics can reveal failures obscured by aggregate performance~\cite{zeng2026workflows}. The broader challenge is to connect such evidence across the complete training process. Similar changes in reward or entropy can arise from different causes, while a fault in one component may become visible only through its effects on another. Detecting an abnormal trajectory therefore does not necessarily establish which component should be changed.

Future research should develop a unified view of training state that connects learning statistics with generated trajectories, evaluator outputs, environment events, and infrastructure behavior. These observations should retain their associated model versions and configuration changes so that symptoms can be traced across stages and over time. Such a view would support diagnosis that distinguishes normal learning transitions from failures and identifies the evidence needed before intervention. Its evaluation should examine diagnostic accuracy, detection delay, and the effect of interventions on subsequent learning, while accounting for the cost of collecting and retaining detailed observations.

\subsection{Cross-Stage Coordination}
\label{sec:cross-stage-coordination}

The techniques reviewed in this survey act on a coupled training process, so their benefits cannot be assumed to accumulate independently. For example, increasing rollout concurrency can improve utilization while changing experience freshness, and revising a reward evaluator can alter both update dynamics and the interpretation of evaluation results. Existing frameworks coordinate distributed training operations~\cite{sheng2025hybridflow,fu2026areal}, while EvoTrainer explores adapting training harnesses using rollout evidence and backtested interventions~\cite{chen2026evotrainer}. These efforts establish important foundations, but leave a broader composition problem: how should sampling, feedback, optimization, recovery, and evaluation controls work together when their objectives or assumptions conflict?

A promising direction is to develop a common coordination layer through which operational methods expose their observations, intervention conditions, and expected effects. This would allow local controllers to account for dependencies, avoid incompatible changes, and reassess decisions as training conditions evolve. For example, a proposed reward correction could trigger revalidation of affected trajectories and checkpoints before further updates proceed. Research should assess combinations of techniques under changing workloads and interacting failures, measuring validated capability gains relative to total time and resource cost. Such evidence is needed to determine when local improvements strengthen the complete training process and when their interactions undermine it.

\subsection{Generalizable RLOps}
\label{sec:generalizable-rlops}

Generalizable RLOps concerns whether diagnostic knowledge and management policies remain effective across models, tasks, and execution settings. Existing approaches provide partial foundations for this goal. RFT-FM detects failures relative to calibrated profiles of normal training behavior~\cite{zhang2026failure}, while Wang et al.~\cite{wang2026moeproxy} construct smaller MoE proxies that reproduce selected failures of larger models. These approaches make operational knowledge reusable within the training conditions or structural relationships they characterize.

Their applicability nevertheless depends on how closely a new setting preserves those reference conditions. A calibrated failure signature may change with the reward distribution or training algorithm, while a proxy deliberately preserving model structure does not establish transfer across substantially different architectures. The general limitation is that evidence of diagnostic effectiveness under specified conditions does not yet provide a reliable criterion for deciding when that knowledge remains applicable elsewhere. Consequently, transferring a monitoring rule or corrective action can require substantial revalidation, and a previously successful intervention may become ineffective or harmful.

Future research should make the applicability of operational knowledge an explicit part of RLOps methods. Failure descriptions and intervention policies should record the conditions supporting their use and distinguish reusable mechanisms from settings that require recalibration. Low-cost pilot runs could then test uncertain transfers before deployment at scale. Evaluation should hold out model families, tasks, reward mechanisms, and execution platforms, measuring both adaptation effort and the reliability of subsequent interventions. This would support accumulated operational experience whose reuse is guided by evidence about its limits.

\subsection{Multi-Run Management}
\label{sec:multi-run-management}

LLM development involves multiple training runs competing for shared resources. MARLaaS and PlexRL improve execution efficiency by coordinating concurrent RL workloads~\cite{yu2026marlaas,zhang2026plexrl}. EvalStop incorporates learning outcomes into platform management by terminating runs after sustained evaluation decline, retaining their best checkpoints, and releasing resources, with scheduling benefits evaluated in simulation~\cite{zhang2026evalstop}. Together, these studies establish mechanisms for sharing infrastructure and avoiding some forms of unproductive computation.

However, the management decisions they support remain tied to particular indicators of progress: execution demand and waiting time in the former systems, and detected quality decline in the latter. These indicators provide limited guidance for allocating additional computation among several healthy runs with uncertain future gains. Efficiently executing all submitted work does not establish whether the available budget is being spent on the most valuable training opportunities. Moreover, allocations can themselves change learning behavior through delays and experience staleness. This leaves an incomplete connection between cluster scheduling and the evolving value of training effort.

Future work should connect resource allocation with estimates of expected improvement, uncertainty, and the cost of interruption or recovery. Such estimates must respect different task objectives instead of directly comparing raw reward values across runs. Management policies should also account for relationships among runs, such as alternative configurations of the same experiment, when deciding which branches to continue or pause. Evaluation should assess the quality of retained models and wasted computation alongside utilization and completion time, under heterogeneous workloads and noisy learning signals. This would extend multi-run management toward productive allocation of training effort across a shared platform.

\subsection{RLOps for RSI}
\label{sec:rlops-for-rsi}

Recursive self-improvement extends operational management to the mechanisms that determine how subsequent improvement is conducted. Existing co-evolution systems provide concrete starting points. VACE validates harness revisions before they affect subsequent RL training, while SafeEvolve combines policy optimization with auditable and reversible harness updates~\cite{qi2026vace,mao2026safeevolve}. These methods demonstrate how model and harness changes can be coordinated within a specified improvement procedure.

A shared boundary is that the search, update, and evaluation procedures still provide an externally specified basis for accepting changes. VACE acknowledges adaptive selection effects from repeated validation reuse, while SafeEvolve identifies its fixed search and update schedules as a restriction on deeper recursive improvement. Their results therefore do not establish the reliability of an improvement process that also revises its evaluators or update rules. Under such changes, the evidence used to accept one generation may lose validity for its successors, while feedback errors and weakened constraints can accumulate across generations~\cite{pu2026trustworthy}. The central challenge is to preserve trustworthy improvement as the improvement process itself evolves.

Future RLOps research should distinguish changes to trained artifacts from changes to the procedures that generate and assess them. Revisions to evaluators or update strategies require independent validation and explicit limits on what the evolving system may modify. Version histories should connect each change to its supporting evidence and inherited effects, enabling recovery of a consistent training configuration when regressions emerge. Longitudinal evaluations should examine sustained gains, capability retention, and the reliability of acceptance decisions across successive generations. This would provide evidence for extending bounded co-evolution toward more autonomous improvement while preserving effective oversight.

\section{Conclusion}
\label{sec:conclusion}

The growing use of reinforcement learning in LLM post-training creates an increasing need for efficient, reliable, and reproducible training operations. In this survey, we provide a systematic overview of LLM RLOps methods throughout the training lifecycle. We organize existing approaches into five stages: Training Setup, Experience Collection, Reward Computation, Policy Optimization, and Model Evaluation. Within each stage, we examine operational concerns, representative management techniques, and evaluation evidence, bringing together research on training infrastructure, feedback reliability, learning stability, and model assessment.

Despite recent progress, important challenges remain in extending individual techniques into dependable operational capabilities. Within a training run, observability must connect evidence across components, while cross-stage coordination must account for dependencies among interventions. Across runs, generalizable RLOps requires establishing when operational knowledge remains applicable, and multi-run management must connect resource allocation with the evolving value of training effort. RLOps for recursive self-improvement further requires preserving reliable evaluation and effective oversight as improvement procedures themselves evolve. Existing studies provide foundations for these directions, but their applicability and combined effectiveness require broader validation.

Advancing LLM RLOps can help reduce wasted computation, improve the dependability of training outcomes, and make operational experience more reusable. As model development becomes increasingly automated, progress will depend on connecting management decisions to validated improvements and preserving control over successive training iterations. We hope this survey provides a foundation for developing more efficient, reliable, and controllable LLM reinforcement learning processes.

\section*{Acknowledgments}

This work is supported by Key RD Project of Guangdong Province
(No.2020B010164003).

\bibliographystyle{ACM-Reference-Format}
\bibliography{sample-base}

\end{document}